\documentclass[12pt]{article}
\usepackage{graphics,cite,amssymb,epsfig,float,psfrag,axodraw}

\newcommand{\lsim}{\raisebox{-0.13cm}{~\shortstack{$<$ \\[-0.07cm] $\sim$}}~}

\newcommand{\defsum}{\raisebox{-0.13cm}{~\shortstack{\Large{$\Sigma$} 
\\[-0.07cm] ${}_{m}$}}~}
\newcommand{\defsumg}{\raisebox{-0.13cm}{~\shortstack{\Large{$\Sigma$} 
\\[-0.07cm] ${}_{m, G}$}}~}

\newcommand{\ra}{\rightarrow}
\newcommand{\ee}{e^+e^-}
\newcommand{\s}{\smallskip}

\newcommand{\nn}{\noindent}
\newcommand{\non}{\nonumber}
\newcommand{\beq}{\begin{eqnarray}}
\newcommand{\eeq}{\end{eqnarray}}
\newcommand{\bev}{\begin{verbatim}}
\newcommand{\eev}{\end{verbatim}}
\newcommand{\tb}{\tan\beta}
\newcommand{\sus}{{\tt SuSpect}}
\def\spheno{{\tt SPheno}}
\def\suspect{{\tt SuSpect}}
\def\softsusy{{\tt SoftSusy}}

\def\be{\begin{equation}}
\def\ee{\end{equation}}
\def\polemz{M_{Z}}
\def\mz{m_{Z}}
\def\drbar{\overline{\rm DR}}
\def\msbar{\overline{\rm MS}}
\def\as{\alpha_s}
\def\sq2{\sqrt{2}}

\def\polemt{M_t}
\def\ov{\overline}

\begin{document}
\baselineskip=16.5pt
\renewcommand{\thefootnote}{*}

\begin{flushright}
PM--02--39\\
CERN TH/2002--325\\
November 2002\\
(updated April 2005)
\end{flushright}

\vspace{.5cm}

\begin{center}

{\large\sc {\bf {\Large \tt  SuSpect}: a Fortran Code for the Supersymmetric}} 

\vspace*{3mm}

{\large\sc {\bf and Higgs Particle Spectrum in the MSSM\footnote{The program
with all relevant information can be downloaded from the web at the http site: 
{\tt www.lpta.univ-montp2.fr/\~\,kneur/Suspect} or  obtained by sending an
E--mail to one of the authors, {\tt abdelhak.djouadi@cern.ch, 
kneur@lpta.univ-montp2.fr, moultaka@lpta.univ-montp2.fr}.}} }

\vspace*{9mm}
\setcounter{footnote}{0}
\renewcommand{\thefootnote}{\arabic{footnote}}

\mbox{{\sc Abdelhak DJOUADI}$^{1,2}$, {\sc Jean--Lo\"{\i}c KNEUR}$^1$ 
and {\sc Gilbert MOULTAKA}$^1$}

\vspace*{0.9cm}

{\it $^1$ Laboratoire de Physique Th\'eorique et Astroparticules, 
UMR5207--CNRS,\\ 
Universit\'e de Montpellier II, F--34095 Montpellier Cedex 5, France.} \s

{\it $^2$ Laboratoire de Physique Th\'eorique d'Orsay, UMR8627--CNRS,\\
Universit\'e Paris--Sud, B\^at. 210, F--91405 Orsay Cedex, France.}

\end{center}

\vspace{1cm}

\begin{abstract}

\nn We present the {\sc Fortran} code {\tt SuSpect} version 2.3, which
calculates the Supersymmetric and Higgs particle spectrum in the Minimal
Supersymmetric Standard Model (MSSM). The calculation can be performed in
constrained models with universal boundary conditions at high scales such as
the gravity (mSUGRA), anomaly (AMSB) or gauge (GMSB) mediated supersymmetry
breaking models, but also in the non--universal MSSM case with R--parity and CP
conservation.  Care has been taken to treat important features such as the
renormalization group evolution of parameters between low and high energy
scales, the consistent implementation of radiative electroweak symmetry
breaking and the calculation of the physical masses of the Higgs  bosons and
supersymmetric particles taking into account the dominant radiative
corrections. Some checks of important theoretical and experimental features,
such as the absence of non desired minima, large fine--tuning in the
electroweak symmetry breaking condition, as well as agreement with precision
measurements can be performed.  The program is simple to use, self--contained
and can easily be linked to other codes; it is rather fast and flexible, thus
allowing scans of the parameter space with several possible options and choices
for model assumptions and approximations.  \end{abstract}

\newpage

\setcounter{footnote}{0}
\renewcommand{\thefootnote}{\arabic{footnote}}

\section*{1. Introduction} 

Supersymmetric theories (SUSY) \cite{SUSY}, which provide an elegant way to
stabilize the large hierarchy between the Grand Unification (GUT) and the
electroweak scales and to cancel the quadratic divergences of the radiative
corrections to the Higgs boson masses, are by far the most studied extensions
of the Standard Model (SM). The most economical low--energy SUSY extension of
the SM, the Minimal Supersymmetric Standard Model (MSSM), which allows for a
consistent unification of the SM gauge couplings  and provides a natural
solution of the Dark Matter problem, has been widely investigated; for reviews
see Refs.~[2-5]. As a corollary, the search for  Supersymmetric particles and
for the extended Higgs spectrum  has become the main goal of present and future
high--energy colliders  \cite{searches}.  \s

It is well--known that in the unconstrained MSSM, it is a rather tedious task
to deal with the basic parameters of the Lagrangian and to derive in an
exhaustive manner their relationship with the physical parameters, i.e. the
particle masses and couplings. This is mainly due to the fact that in  the
MSSM, despite of the  minimal gauge group, minimal particle content, minimal
couplings imposed by R--parity conservation and the minimal set of soft
SUSY-breaking parameters, there are more than a hundred new parameters
\cite{parameters}. Even if one constrains the model to have a viable
phenomenology [we will call later such a model the phenomenological MSSM],
assuming for instance no intergenerational mixing to avoid flavor changing
neutral currents, no  new source of CP violation, universality of first and
second generation  sfermions to cope with constraints from kaon physics, etc..,
there are still more than 20 free parameters left.  This large number of input
enters in the evaluation of the masses of ${\cal O}(30)$ SUSY particles and
Higgs bosons as well as their complicated couplings, which involve several
non--trivial aspects, such as the mixing between different states, the Majorana
nature of some particles, etc.  The situation becomes particularly difficult if
one aims at rather precise calculations and hence, attempts to include some
refinements such as higher order corrections, which for the calculation of a
single parameter need the knowledge of a large part of, if not the whole,
spectrum. \s

Thus, the large number of free parameters in the unconstrained or even
phenomenological MSSM, makes a detailed phenomenological analysis of the
spectra and the comparison with the outcome or expectation from experiment, a
daunting task, if possible at all. Fortunately, there are well motivated
theoretical models where the soft SUSY--breaking parameters obey a number of
universal boundary conditions at the high (GUT) scale, leading to only a
handful of basic parameters. This is the case for instance of the minimal
Supergravity model (mSUGRA) \cite{mSUGRA}, where it is assumed that
SUSY--breaking occurs in a hidden sector which communicates with the visible
sector only through ``flavor--blind'' gravitational interactions. This leads to
the simpler situation where the entire spectrum of superparticles and Higgs
bosons is determined by the values of only five free parameters and makes
comprehensive scans of the parameter space and detailed studies of the spectrum
feasible.\s

However, there are also similarly constrained and highly predictive alternative
SUSY--breaking models in the literature, such as anomaly mediated \cite{AMSB,
AMSB1, AMSBp} or gauge mediated \cite{GMSB,GMSBp} SUSY--breaking models for
instance, which should be investigated as well. We then have to trade a
complicated situation where we have one model with many input parameters, with
a not less complicated situation where we have many models with a small number
of basic parameters. In addition, in these unified models, the low--energy
parameters are derived from the high--energy (GUT and/or possibly some
intermediate scales) input parameters through Renormalization Group Equations
(RGE) and they should also necessarily involve radiative electroweak symmetry
breaking (EWSB), which sets additional constraints on the model.  The
implementation of the RG evolution and the EWSB mechanism poses numerous
non--trivial technical problems if they have to be done in an accurate way,
i.e. including higher order effects. This complication has to be added to the
one from the calculation of the particle masses and couplings with radiative
corrections (RC) which is still present.\s

Therefore, to deal with the supersymmetric spectrum in all possible cases, one
needs very sophisticated programs to encode all the information and,
eventually, to pass it to other programs or Monte Carlo generators to simulate
the physical properties of the new particles, decay branching ratios,
production cross sections at various colliders, etc... These programs should
have a high degree of flexibility in the choice of the model and/or the input
parameters and an adequate level of approximation at different stages, for
instance in the incorporation of the RGEs, the handling of the EWSB  and the
inclusion of radiative corrections to (super)particle masses, which in many
cases can be very important. They should also be reliable, quite fast to allow
for rapid comprehensive scans of the parameter space and simple enough to be
linked with  other programs.  There are several public codes, in particular
{\tt ISASUGRA} \cite{ISASUGRA}, {\tt SOFTSUSY} \cite{SOFTSUSY} and {\tt SPHENO}
\cite{SPHENO}, as well as a number of private codes, which deal with this
problem.  In this paper we present our program \sus. \s

{\tt SuSpect}, in its latest version 2.3 that we present here, is a {\sc
Fortran} code which calculates the supersymmetric and Higgs particle spectrum
in the constrained  and unconstrained MSSMs. The acronym is an abbreviation of
{\tt Su}sy{\tt Spect}rum and successive previous public versions of the code
were available starting from 1997 and have been  described in
Ref.~\cite{SUSPECT}. At the present stage, it deals with the ``phenomenological
MSSM" with 22 free parameters defined either at a low or high energy scale,
with the  possibility of RG evolution to arbitrary scales, and the most studied
constrained models, namely mSUGRA, AMSB and GMSB.  Many ``intermediate" models 
[e.g. constrained models but without unification of gaugino or scalar masses, 
etc..] are easily  handled. The program includes the three major
ingredients  which should be incorporated in any algorithm for the constrained
MSSMs: $i)$ renormalization group evolution of parameters between a low
energy scale [e.g. $M_Z$ and/or the EWSB scale] and a
high--energy scale [e.g. $M_{\rm GUT}$], [17--19];  $ii)$ consistent
implementation of radiative electroweak symmetry breaking [loop corrections to
the effective potential are included using the tadpole method] [20--23];
$iii)$ calculation of the physical (pole) masses of the superparticles and
Higgs bosons, including all relevant features such as the mixing between
various states  and the radiative corrections when important [24--34]. \s 

The present version includes new options to read input files and write output
files in the recently adopted format of the {\em SUSY Les Houches Accord}
(SLHA) for interfacing the spectrum generators with other computer codes
\cite{slha}.  The code contains three types of source files: $i$) the main
subroutine {\tt suspect2.f} together with a separate routine {\tt
twoloophiggs.f} for the two--loop radiative corrections in the Higgs sector,
$ii)$ a separate calling routine file {\tt suspect2$\_$call.f}, and $iii)$ two
possible input files {\tt suspect2.in} (in the original \sus\ format) or {\tt
suspect2$\_$lha.in} (in the SLHA format).  Any choice and option is driven
either from one of the two input files [which is sufficient and convenient when
dealing with a few model points] or alternatively directly from the {\tt
suspect2$\_$call.f} file, which also provides examples of calls for different
model choices with all the necessary features [this option being useful when
interfacing with other routines or when performing scans over the parameter
space].  The program has several flags which allow to select the model to be
studied and its input parameters, the level of accuracy of the algorithm [e.g. 
the iterations for the RGEs and the convergence of the EWSB], the level of
approximation in the calculation of the various (s)particle masses [e.g. 
inclusion or not of RC].   Besides the fact that it is  flexible, the code is
self--contained [the default version includes all routines needed for the
calculation], rather fast [thus allowing large scans of the parameter space]
and can be easily linked to other routines or Monte--Carlo generators [e.g.\,to
calculate branching ratios, cross sections, relic densities]. All results,
including comments when useful and some theoretical and experimental
constraints, are found in the output file {\tt suspect2.out} (in the original
\sus\ format by default) or alternatively in the  output file {\tt
suspect2$\_$lha.out}, which are created at any run of the program. It is hoped
that  the code may be readily usable even without much prior knowledge of the
MSSM.\s

This ``users' manual" for the program, is organized as follows. In section 2,
we briefly discuss the main ingredients of the unconstrained and
phenomenological  MSSMs as well as the constrained models mSUGRA, AMSB and
GMSB, to set the  notations and conventions used in the program.  In section 3,
we summarize the procedure for the calculation of the (s)particle spectrum: the
soft SUSY--breaking terms [including the treatment of the input, the RG
evolution and the implementation of EWSB], the physical particle masses
[summarizing our conventions for the sfermion, gaugino and Higgs sectors].  We
also discuss the theoretical [CCB, UFB, fine--tuning] and experimental
[electroweak precision observables, muon $g-2$, $b \to s \gamma$ branching
fraction] constraints which can be imposed on the spectra, and how these are
implemented in the code.  In section 4, we summarize the basic  practical 
facts about the program  and discuss the content of the input and  output files
with the possible choices; we then make a brief comparison with other existing
codes, discuss the interface with other programs and how the program is
maintained on the web. A conclusion will be given in section 5. In the

\section*{2. The constrained and unconstrained MSSMs}

In this section, we summarize the basic assumptions which define the MSSM and
the various constraints which can be imposed on it. This will also set the
notations and conventions used in the program. We mainly focus on the
unconstrained MSSM, the phenomenological MSSM with 22 free parameters,  as well
as on constrained models such as the minimal Supergravity (mSUGRA), anomaly
mediated  (AMSB) and gauge mediated (GMSB) supersymmetry breaking models.

\subsection*{2.1 The unconstrained MSSM} 

The unconstrained MSSM is defined usually by the following four basic assumptions
\cite{MSSMdef}:\s

\nn {\it (a) Minimal gauge group:}  the MSSM is based on the group  ${\rm
SU(3)_C \times  SU(2)_L \times  U(1)_Y}$, i.e. the SM symmetry. SUSY implies
then that the spin--1 gauge bosons and their spin--1/2 partners, the gauginos
[bino $\tilde{B}$, winos $\tilde{W}_{1-3}$ and gluinos $\tilde{G}_{1-8}$], are
in vector  supermultiplets.\s

\nn {\it (b) Minimal particle content:} there are only three generations of
spin--1/2 quarks and leptons as in the SM. The
left-- and right--handed chiral fields belong to chiral superfields together
with their spin--0 SUSY partners, the squarks and sleptons: ${\hat{ Q}}, {\hat{
u}}_{R}, {\hat{ d}}_{R}, {\hat{ L}},  {\hat{ l}}_{R}$. In addition, two chiral
superfields $\hat{H}_d$, $\hat{H}_u$ with respective hypercharges $-1$ and $+1$
for the cancellation of chiral anomalies, are needed. Their scalar components,
$H_d$ and $H_u$, give separately masses to the isospin +1/2 and $-$1/2 fermions
and lead to five Higgs particles: two CP--even $h,H$ bosons, a pseudoscalar $A$
boson  and two charged $H^\pm$ bosons. Their spin--1/2 superpartners, the
higgsinos, will mix with the winos and the bino, to give the ``ino" mass 
eigenstates: the two charginos $\chi_{1,2}^\pm$ and the four neutralinos
$\chi^0_{1,2,3,4}$. \s

\nn {\it (c) Minimal Yukawa interactions and R--parity conservation:} to
enforce lepton  and baryon  number conservation, a discrete and multiplicative
symmetry called R--parity is imposed.  It is defined by $R_p= (-1)^{2s+3B+L}$,
where L and B are the lepton and baryon numbers and $s$ is the spin quantum
number. The R--parity quantum numbers are then $R_p=+1$ for the ordinary
particles [fermions, gauge and Higgs bosons], and $R_p=-1$ for their
supersymmetric partners. In practice, the conservation of $R$--parity has
important consequences: the SUSY particles are always produced in pairs, in
their decay products there is always an odd number of SUSY particles, and the
lightest SUSY particle (LSP) is absolutely stable.  \s

The three conditions listed above are sufficient to completely determine  a
globally supersymmetric Lagrangian. The kinetic part of the Lagrangian is
obtained by generalizing the notion of covariant derivative to the SUSY case.
The most general superpotential, compatible with gauge invariance, 
renormalizability and R--parity conservation is written as: 
\begin{equation}
W=\sum_{i,j=gen} - Y^u_{ij} \, {\hat {u}}_{Ri} \hat{H_u}.{\hat{ Q}}_j+
     Y^d_{ij} 
\, {\hat{ d}}_{Ri} \hat{H}_d.{\hat{ Q}}_j+
       Y^l_{ij} \,{\hat{ l}}_{Ri} \hat{H}_u.{\hat{ L}}_j+
     \mu \hat{H}_u.\hat{H}_d
\label{defW}
\end{equation}
The product between SU(2)$_{\rm L}$ doublets reads $H.Q \equiv \epsilon_{a b}
H^a Q^b$ where $a, b$ are SU(2)$_{\rm L}$ indices and $ \epsilon_{12}=1 = -
\epsilon_{21}$, and $Y^{u,d,l}_{ij}$ denote the Yukawa couplings among
generations. The first three terms in the previous expression are nothing else
but a superspace generalization of the Yukawa interaction in the SM, while the 
last term is a globally supersymmetric Higgs mass term.  The supersymmetric 
part of the tree--level potential $V_{\rm tree}$ is the sum of the  so--called
F-- and D--terms \cite{DF}, where the F--terms come from the superpotential 
through
derivatives with respect to all scalar fields  $\phi_{a}$, $V_{F}={\sum_{a} 
|W^{a}|^2}$ with $W^{a} = \partial{W}/\partial{ \phi_a}$, and the D--terms 
corresponding to the ${\rm U(1)_Y}$, ${\rm SU(2)_L}$, and ${\rm  SU(3)_C}$
gauge symmetries  are given by $V_{D}= \frac{1}{2}  \sum_{i=1}^{3}  (\sum_{a}
g_i \phi_a^* T^i \phi_a)^2$ with $T^i$ and $g_i$ being the generators  and the
coupling constants of the corresponding gauge groups. \bigskip

\nn {\it (d) Minimal set of soft SUSY--breaking terms:} 
to break Supersymmetry, while preventing the reappearance of the quadratic
divergences [soft breaking], one adds to the supersymmetric Lagrangian a set
of terms which explicitly but softly break SUSY \cite{soft}: 
\begin{itemize} 
\item[$\bullet$] Mass terms for the gluinos, winos and binos:
\beq
- {\cal L}_{\rm gaugino}=\frac{1}{2} \left[ M_1 \tilde{B}  
\tilde{B}+M_2 \sum_{a=1}^3 \tilde{W}^a \tilde{W}_a +
M_3 \sum_{a=1}^8 \tilde{G}^a \tilde{G}_a  \ + \ {\rm h.c.} 
\right]
\eeq
\item[$\bullet$] Mass terms for the scalar fermions: 
\beq
-{\cal L}_{\rm sfermions} = 
{\sum_{i=gen} m^2_{{\tilde {Q}}i} {\tilde{Q}}_i^{\dagger}{\tilde{Q}}_i+
m^2_{{\tilde{ L}}i} {\tilde{L}}_i^{\dagger} {\tilde{L}}_i +
         m^2_{ {\tilde{u}}i} |{\tilde{u}}_{R_i}|^2+m^2_{ {\tilde{d}}i} 
|{\tilde{d}}_{R_i}|^2+  m^2_{{\tilde{l}}i} | {\tilde{l}}_{R_i}|^2}   
\eeq
\item[$\bullet$] Mass and bilinear terms for the Higgs bosons: 
\beq
-{\cal L}_{\rm Higgs} = m^2_{H_u} H_u^{\dagger} H_u+m^2_{H_d}  H_d^{\dagger} 
H_d + B \mu (H_u.H_d + {\rm h.c.} ) 
\eeq
\item[$\bullet$] Trilinear couplings between sfermions and Higgs bosons 
\beq
-{\cal L}_{\rm tril.}= 
{\sum_{i,j=gen} { \left[ A^u_{ij} Y^u_{ij}  {\tilde{u}}_{R_i} H_u. 
{\tilde{Q}}_j+
A^d_{ij} Y^d_{ij}  {\tilde{d}}_{R_i} H_d.{\tilde{Q}}_j
+A^l_{ij} Y^l_{ij} {\tilde{l}}_{R_i} H_u.{\tilde{L}}_j\ + \ {\rm h.c.} 
\right] }}
\eeq
\end{itemize} 
The soft SUSY--breaking scalar potential is the sum of the three last terms:
\beq
V_{\rm soft} = -{\cal L}_{\rm sfermions} -{\cal L}_{\rm Higgs}-
{\cal L}_{\rm tril.}
\eeq
Up to now, no constraint is applied to this Lagrangian, although for generic
values of the parameters, it might lead to severe phenomenological problems 
such as flavor changing neutral currents [FCNC] and unacceptable amount 
of additional CP--violation \cite{flavorev}, color and/or charge breaking minima
\cite{CCBold}, an incorrect value of the $Z$ boson mass, etc... The MSSM 
defined by the four hypotheses $(a)$--$(d)$ above, will be called the 
unconstrained MSSM. 

\subsection*{2.2 The ``phenomenological" MSSM}

In the unconstrained MSSM, and in the general case where one allows for
intergenerational mixing and complex phases, the soft SUSY breaking terms will
introduce a huge number (105) of unknown parameters, in addition to the 19
parameters of the SM \cite{parameters}. This large number of free parameters
makes any phenomenological analysis in the general MSSM very complicated as
mentioned previously. In addition, many ``generic'' sets of these parameters
are excluded by the severe phenomenological constraints discussed above. A
phenomenologically viable MSSM can be defined by making the following three
assumptions: $(i)$ All the soft SUSY--breaking parameters are real and
therefore there is no new source of CP--violation generated, in addition to the
one from the CKM matrix.  $(ii)$ The matrices for the sfermion masses and for
the trilinear couplings are all diagonal,  implying the absence of FCNCs at the
tree--level.  $(iii)$ First and second sfermion generation universality at low
energy to  cope with the severe constraints from $K^0$--$\bar{K}^0$ mixing, etc
[this is also motivated by the fact that one can neglect for simplicity all the
masses of the first and second generation fermions which are too small  to
have any effect on the running of the SUSY--breaking parameters].\s 

Making these three assumptions will lead to 22 input parameters only: 

\nn \hspace*{2cm} $\tan \beta$: the ratio of the vevs of the two--Higgs doublet
fields.\\
 \hspace*{2cm} $m^2_{H_u}, m^2_{H_d}$: the Higgs mass parameters squared. \\
 \hspace*{2cm} $M_1, M_2, M_3$: the bino, wino and gluino mass parameters. \\
 \hspace*{2cm} $m_{\tilde{q}}, m_{\tilde{u}_R}, m_{\tilde{d}_R}, 
               m_{\tilde{l}}, m_{\tilde{e}_R}$: the first/second generation
 sfermion mass parameters.\\ 
  \hspace*{2cm} $m_{\tilde{Q}}, m_{\tilde{t}_R}, m_{\tilde{b}_R}, 
               m_{\tilde{L}}, m_{\tilde{\tau}_R}$: the third generation
 sfermion mass parameters.\\
  \hspace*{2cm} $A_u, A_d, A_e$: the first/second generation trilinear 
  couplings. \\
  \hspace*{2cm} $A_t, A_b, A_\tau$: the third generation trilinear couplings. \s

Two remarks can be made at this stage: 
$(i)$ The Higgs--higgsino (supersymmetric) mass parameter $|\mu|$ (up to a
sign) and the soft SUSY--breaking  bilinear Higgs term $B$ are determined,
given the above parameters,  
through the electroweak symmetry  breaking conditions as will be discussed
later. Alternatively, one can trade  the values of $m^2_{H_u}$ and $m^2_{H_d}$
with the ``more physical" pseudoscalar Higgs boson  mass $M_A$ and parameter
$\mu$ [such an alternative choice is explicitly possible in \sus\ by
appropriate setting of the input parameters]. 
$(ii)$ Since the trilinear sfermion couplings will be always multiplied by the
fermion masses, they are important only in the case of the third generation.
However, there are a few (low scale) situations, such as the  muon $(g-2)$ and
the neutralino--nucleon scattering for direct Dark Matter  searches, where they
will play a role. We thus consider them as input.\s

Such a model, with this relatively moderate number of parameters [especially
that, in general, only a small subset contributes dominantly 
(i.e. at tree-level) when one looks at a given sector
of the model] has much more predictability and is much easier to investigate
phenomenologically, compared to the unconstrained MSSM. We will refer to this 
22 free input parameters model as the  ``phenomenological" MSSM or pMSSM
\cite{GDR}. 

\subsection*{2.3 The mSUGRA model}

Almost all problems of the general or unconstrained MSSM are solved at once if
the soft SUSY--breaking parameters obey a set of universal boundary conditions
at the GUT scale. If one takes these parameters to be real, this solves all
potential problems with CP violation as well. The underlying assumption is that
SUSY--breaking occurs in a hidden sector which communicates with the visible
sector only through gravitational--strength interactions, as specified by
Supergravity.  Universal soft breaking terms then emerge if these Supergravity
interactions are ``flavor--blind'' [like ordinary gravitational interactions]. 
This is assumed to be the case in the constrained MSSM or minimal Supergravity
(mSUGRA) model \cite{mSUGRA}. \s

Besides the unification of the gauge coupling constants $g_{1,2,3}$ of the
U(1), SU(2) and SU(3) groups, which is well verified given the experimental
results
from LEP1 \cite{PDG} and which can be viewed as fixing the Grand Unification
scale $M_{\rm GUT} \sim 2 \cdot 10^{16}$ GeV \cite{LEPunif}, the unification
conditions in mSUGRA, are as  follows: \s

-- Unification of the gaugino [bino, wino and gluino] masses: 
\beq
M_1 (M_{\rm GUT})=M_2(M_{\rm GUT})=M_3(M_{\rm GUT}) \equiv m_{1/2}
\eeq

-- Universal scalar [i.e. sfermion and Higgs boson] masses [$i$ is the 
generation index]: 
\beq
M_{\tilde{Q}_i} (M_{\rm GUT}) &=& M_{\tilde{u}_{Ri}} (M_{\rm GUT})
= M_{\tilde{d}_{Ri}}(M_{\rm GUT})  =M_{\tilde{L}_i} (M_{\rm GUT}) 
= M_{\tilde{l}_{Ri}} (M_{\rm GUT}) \non \\
&=& M_{H_u}(M_{\rm GUT}) =M_{H_d} (M_{\rm GUT}) \equiv  m_0
\eeq

-- Universal trilinear couplings: 
\beq
A^u_{ij} (M_{\rm GUT}) = A^d_{ij} (M_{\rm GUT}) = A^l_{ij} (M_{\rm
GUT}) \equiv  A_0 \, \delta_{ij}
\eeq

Besides the three parameters $m_{1/2}, m_0$ and $A_0$, the supersymmetric
sector is described at the GUT scale by the bilinear coupling $B$ and the
supersymmetric Higgs(ino) mass parameter $\mu$. However, one has to require
that EWSB takes place at some low energy scale. This results in two necessary
minimization conditions of the two--Higgs doublet scalar potential which, at
the tree--level, has the form [to have a more precise description, one--loop
corrections to the scalar potential have to be included, as will be discussed
later]: %
\begin{eqnarray} 
V_{\rm Higgs} &=& \overline{m}_1^2 H_d^{\dagger} H_d + \overline{m}_2^2 
H_u^{\dagger} H_u + \overline{m}_3^2 (H_u \cdot H_d + {\rm h.c.}) \nonumber \\
 &+& \frac{g_1^2+g_2^2}{8}  (H_d^{\dagger} H_d - H_u^{\dagger} H_u)^2 + 
\frac{g_2^2}{2} (H_d^{\dagger}  H_u) (H_u^{\dagger}  H_d)
\label{vhiggs}, 
\end{eqnarray}
where we have used the usual short--hand notation: $\overline{m}_1^2= m^2_{H_d}
+\mu ^2 ,  \overline{m}_2^2= m^2_{H_u}+\mu ^2$, $\overline{m}_3^2= B\mu$ and
the SU(2) invariant product of the two doublets $\phi_1 \cdot \phi_2 = 
\phi_1^1 \phi_2^2 - \phi_1^2 \phi_2^1$. The two minimization equations
$\partial V_{\rm Higgs} / \partial H_d^0 = \partial V_{\rm Higgs} /
\partial H_u^0 = 0$ can be solved for $\mu^2$ and $B \mu$:
\begin{eqnarray} \label{eq:ewsb}
\mu^2 &=& \frac{1}{2} \bigg[ \tan 2\beta (m^2_{H_u} \tan \beta
- m^2_{H_d} \cot \beta) -M_Z^2 \bigg] \non \\
B\mu &=& \frac{1}{2} \sin 2\beta \Bigg[ m^2_{H_u} + m^2_{H_d} + 2
\mu^2 \Bigg] 
\end{eqnarray}
(which are of course valid
as well in the more general unconstrained MSSM). 
Here, $M_Z^2=(g_1^2+g_2^2) \cdot (v_u^2 + v_d^2) /4$ and $\tan \beta= v_u/v_d$
is defined in terms of the vacuum expectation values of the two neutral Higgs
fields. Consistent EWSB is only possible if eq.~(\ref{eq:ewsb}) gives a
positive value of $\mu^2$. The sign of $\mu$ is not determined. Therefore, in
this model, one is left with only four continuous free parameters, and an
unknown sign:
\beq 
\tan \beta \ , \ m_{1/2} \ , \ m_0 \ , \ A_0 \ , \ \ {\rm sign}(\mu). 
\eeq 
All the soft SUSY breaking parameters at the weak scale are then
obtained through Renormalization Group Equations.

\subsection*{2.4 The AMSB model}

In mSUGRA, Supersymmetry is broken in a hidden sector and the breaking is
transmitted to the visible sector by gravitational interactions. In Anomaly
Mediated Supersymmetry Breaking models, the SUSY--breaking occurs also in a
hidden sector, but it is transmitted to the visible sector by the super--Weyl
anomaly \cite{AMSB, AMSB1}. The gaugino, scalar masses and trilinear couplings 
are then simply related to the scale dependence of the gauge and matter kinetic
functions.  This leads to soft SUSY--breaking scalar masses for the first two
generation sfermions that are almost diagonal [when the small Yukawa couplings 
are neglected]. \s
 
In terms of the gravitino mass $m_{3/2}$ [which is much larger than the 
gaugino and squark masses, a cosmologically appealing feature], the  $\beta$
functions for the gauge and Yukawa couplings $g_a$ and $Y_i$, and  the
anomalous dimensions $\gamma_i$ of the chiral superfields, the soft SUSY 
breaking terms are given by:
\beq
M_a &=& \frac{\beta_{g_a}}{g_a} m_{3/2} \ , \  \non \\
A_i &=& \frac{\beta_{Y_i}}{Y_i}  m_{3/2} \non \\  
m_i^2 &=& -\frac{1}{4} \left( \Sigma_a \frac{\partial \gamma_i}{\partial g_a}
\beta_{g_a} +
\Sigma_k \frac{\partial \gamma_i}{\partial Y_k} \beta_{Y_k} \right) m_{3/2}^2 
\eeq
These equations are RG invariant and thus valid at any scale and make the model
highly predictive.  The additional parameters, $\mu^2$ and $B$ are obtained as
usual by requiring the correct breaking of the electroweak symmetry. One then
has, in principle, only three input parameters $m_{3/2}, \tan\beta$ and
sign$(\mu)$. However, this rather simple picture is spoiled by the fact that
the anomaly mediated contribution to the slepton scalar masses squared is
negative and the sleptons are in general tachyonic. This problem can be cured
by adding a positive non--anomaly mediated contribution to the soft masses. The
simplest phenomenological way of parameterizing the non--anomaly contribution
is to add a common mass parameter $m_0$ at the GUT scale, which would be then
an additional input parameter to all the (squared) scalar masses. However in
the general case, the non--anomaly mediated contribution might be different for
different scalar masses and depend on the specific model which has been chosen.
One should then write a general non--anomalous contribution at the GUT scale for
each scalar mass squared:
\beq
m_{\tilde{S}_i}^2 = c_{S_i}m_0^2 -\frac{1}{4} \left( \Sigma_a \frac{\partial 
\gamma_i} {\partial g_a} \beta_{g_a} + \Sigma_k \frac{\partial \gamma_i}
{\partial Y_k} \beta_{Y_k}  \right)  m_{3/2}^2 +{\rm D\, terms.} 
\eeq
where the coefficients $c_{S_i}$ depend on the considered model. \s

A few examples of models with different non--anomalous contributions are: \s

\nn -- The minimal anomaly mediated supersymmetry breaking model with a universal 
$m_0$ \cite{AM1}:
\beq
c_{Q}=c_{u_R}=c_{d_R}=c_{L}=c_{e_R}=c_{H_u}=c_{H_d}=1
\eeq
-- The gaugino assisted AMSB model where one assumes that gauge and gaugino 
fields reside in the bulk of an extra dimension \cite{AM2}:
\beq
c_{Q}=21/10,c_{u_R}=8/5,c_{d_R}=7/5,c_{L}=9/10,c_{e}=3/5,c_{H_u}=9/10c_{H_d} 
\eeq
-- Models where an extra U(1) factor is added; a  particular scenario is
interesting 
phenomenologically since   it leads to a light top squark \cite{AM3}: 
\beq
c_{Q}=3,c_{u_R}=c_{d_R}=-1,c_{L}=c_{e}=1,c_{H_u}=c_{H_d}=-2
\eeq 

A simple way to account for all the different models is to add to the three
continuous and one discrete original basic parameters, the set of coefficients
$c_{S_i}$ as input to specify, and therefore one would have the set of  input 
parameters:
\beq
m_0 \ , \ m_{3/2} \ , \ \tb \ , \ {\rm sign}(\mu) \ {\rm and} \ c_{S_i}
\eeq 
This is the approach that we will follow in the program. 

\subsection*{2.5 The GMSB model}

In Gauge Mediated Supersymmetry Breaking models, SUSY--breaking is
transmitted to the MSSM fields via the SM gauge interactions. In the original
scenario \cite{gmsb1}, the model consists of three distinct sectors: a secluded
sector where SUSY is broken, a ``messenger" sector containing a singlet field
and messenger fields with ${\rm SU(3)_c\times SU(2)_L\times U(1)_Y}$ quantum
numbers, and a sector containing the fields of the MSSM. Another possibility,
the so--called ``direct gauge mediation" \cite{gmsb2} has only two sectors: one
which is responsible for the SUSY breaking and contains the messenger fields,
and another sector consisting of the MSSM fields. In both cases, the soft 
SUSY--breaking masses for the gauginos and squared masses for the sfermions 
arise,
respectively, from one--loop and two--loop diagrams involving the exchange of
the messenger fields, while the trilinear Higgs--sfermion--sfermion couplings
can be taken to be negligibly small at the messenger scale since they are [and
not their square as for the sfermion masses] generated by two--loop gauge
interactions.  This allows an automatic and natural suppression of FCNC and
CP--violation; for a review see, Ref.~\cite{GMSB}. \s

In the GMSB models that we will consider, the source of SUSY breaking is
parameterized by an ${\rm SU(3)_C \times SU(2)_L \times U(1)_Y}$ gauge--singlet
chiral  superfield $\hat{S}$ whose scalar and auxiliary components acquire
vacuum  expectation values denoted by $S$ and $F_S$, respectively.  We assume
$n_{\hat{q}}$ pairs of $\hat{q}, \hat{\bar{q}}$ quark--like [resp.
$n_{\hat{l}}$ pairs of  $\hat{l}, \hat{\bar{l}}$ lepton--like]  messenger
superfields transforming as  $(3, 1, -\frac{1}{3}),  (\bar{3}, 1,
\frac{1}{3})$  [resp. $(1, 2, \frac{1}{2}), (1, 2, -\frac{1}{2})$] under ${\rm
SU(3)_C \times SU(2)_L \times U(1)_Y}$ and coupled to $\hat{S}$ through a
superpotential  of the form $\lambda \hat{S} \hat{q} \hat{ \bar{q}} + \lambda 
\hat{S} \hat{l} \hat{\bar{l}}$. Soft SUSY--breaking parameters are then
generated at the messenger scale  $M_{\rm mes} = \lambda S$,  
\begin{eqnarray}
M_G(M_{\rm mes}) &=& \frac{\alpha_G(M_{\rm mes})}{4 \pi} \Lambda \,
g \bigg(\frac{\Lambda}{M_{\rm mes}} \bigg) \defsum N^G_R(m)
  \label{gasoft} \\
m_s^2(M_{\rm mes}) & =& 2 \Lambda^{2} f\bigg(\frac{\Lambda}{M_{\rm mes}} 
\bigg)\defsumg  \bigg[ \frac{\alpha_G(M_{\rm mes})}{4 \pi} \bigg]^2  
N^G_R(m) C_R^G(s)
\label{scalsoft} \\
A_{f} (M_{\rm mes})  &\simeq & 0
\end{eqnarray}
where $\Lambda=F_S/S$, $G= {\rm U(1), SU(2), SU(3)}$, $m$ labels the 
messengers and $s$  runs over the Higgs doublets  as well as the left--handed 
doublets and right--handed singlets of squarks and sleptons. The 
one-- and two--loop functions $g$ and $f$ are given by [Li$_2$ is the 
Spence function]:
\beq
g(x)&=&  \frac{1}{x^2} [(1+x)\log (1+x)+(1-x) \log (1-x)]  \non \\
f(x) &=& \frac{1+x}{x^2} \bigg[\log (1+x) -2 {\rm Li}_2 \bigg( \frac{x}{1+x} 
\bigg) + \frac{1}{2} {\rm Li}_2 \bigg( \frac{2x}{1+x} \bigg) \bigg] + 
(x \leftrightarrow -x)  
\eeq 
Defining the Dynkin index $N_R^G$ by
\begin{equation}
{\rm Tr}\, (T_R^a T_R^b) = \frac{N_R^G}{2} \delta^{a b}
\end{equation}
\noindent
for non--abelian groups, and $N^{U(1)_Y} = (6/5) Y^2$ where $Y \equiv 
Q_{\rm EM}- T_3$, one has (see eq.(\ref{gasoft}))
\begin{eqnarray}
\defsum N^{U(1)_Y}_R(m) &=& \frac{1}{5} (2 n_{\hat{q}} + 3 n_{\hat{l}}) 
\nonumber \\
\defsum N^{SU(2)_L}_R(m) &=& n_{\hat{l}} \nonumber \\
\defsum N^{SU(3)_c}_R(m) &=& n_{\hat{q}} 
\end{eqnarray}
\noindent
With the Casimir invariant $C_{\bf N}^G$ given by 
\begin{equation}
\Sigma_a T^a_{\bf N} T^a_{\bf N} = C_{\bf N}^{SU(N)} {\bf 1} = 
\frac{N^2-1}{2 N} {\bf 1}
\end{equation}
\noindent
for  the ${\bf N}$  of $SU(N)$, and $C^{U(1)_Y} = (3/5) Y^2$, 
one finds for 
$${\cal N C}(s) \equiv \defsumg \bigg[ \frac{\alpha_G(M_{\rm mes})}{4 \pi}
\bigg]^2  N^G_R(m) C_R^G(s)$$
\noindent
(see eq.(\ref{scalsoft})) the following values:
\begin{eqnarray}
{\cal N C}(\tilde{Q}) &=& \frac{1}{16 \pi^2} \left[ (\frac{n_{\hat{l}}}{100} 
+ \frac{n_{\hat{q}}}{150}) \alpha_1^2 + \frac{3 n_{\hat{l}}}{4} \alpha_2^2 + 
\frac{4 n_{\hat{q}}}{3} \alpha_3^2 \right] \nonumber \\
{\cal N C}(\tilde{U})&=& \frac{1}{16 \pi^2} \left [
(\frac{4 n_{\hat{l}}}{25} + \frac{8 n_{\hat{q}}}{75}) \alpha_1^2 +
\frac{4 n_{\hat{q}}}{3} \alpha_3^2   \right] \nonumber \\
{\cal N C}(\tilde{D})&=& \frac{1}{16 \pi^2} \left[
(\frac{n_{\hat{l}}}{25} + \frac{2 n_{\hat{q}}}{75}) \alpha_1^2 +
\frac{4 n_{\hat{q}}}{3} \alpha_3^2   \right] \nonumber \\
{\cal N C}(\tilde{L})&=& \frac{1}{16 \pi^2} \left[
(\frac{9 n_{\hat{l}}}{100} + \frac{3 n_{\hat{q}}}{50}) \alpha_1^2 +
\frac{3 n_{\hat{l}}}{4} \alpha_2^2 \right] \nonumber \\
{\cal N C}(\tilde{E})&=& \frac{1}{16 \pi^2} \left[
(\frac{9 n_{\hat{l}}}{25} + \frac{6 n_{\hat{q}}}{25}) \alpha_1^2 
\right] \nonumber \\
{\cal N C}(\tilde{H}_u)&=& {\cal N C}(\tilde{H}_d) = {\cal N C}(\tilde{L})
\end{eqnarray} 
The freedom in choosing independently the number of $n_{\hat{q}}$ and
$n_{\hat{l}}$
messengers allows to study various model configurations: for instance when the 
messengers are assumed to form complete representations of some grand 
unification group (e.g.  ${\bf 5} + {\bf \bar{5}}$ of SU(5)) where 
$n_{\hat{q}} = n_{\hat{l}}$, or when they transform under larger unification
group factors with some extra discrete symmetries where typically
$n_{\hat{q}} \neq n_{\hat{l}}$ \cite{barbieri}.
[When $n_{\hat{q}} = n_{\hat{l}} = 1$ one retrieves the 
minimal model \cite{gmsb1}. In this case the gaugino masses  
have the same {\sl relative} values as if they were
unified at $M_{\rm GUT}$ despite the fact the boundary conditions are set at
$M_{\rm mes}$ and that scalar masses are flavor independent.
Furthermore when 
$\Lambda/M_{\rm mes} \ll 1$, one has $f(x) \simeq g(x) \simeq 1$.]   
In addition, some constraints are
in general needed in order to have a viable spectrum, for instance:
$\Lambda/M_{\rm mes} < 1$ to avoid negative mass squared for bosonic members of
the messenger scale and $\Lambda/M_{\rm mes} \lsim 0.9$ to avoid too much
fine--tuning in EWSB. Note also that $n_{\hat{q}} > n_{\hat{l}}$
improves the fine--tuning issue \cite{agashe}.\s

\noindent
Once the boundary conditions are set at $M_{\rm mes}$, the low energy
parameters are obtained via the usual RGEs and the proper breaking of
the EW symmetry is required. 

Therefore, in the GMSB model that we are considering, there are six input 
parameters
\beq
\tan \beta \ , \ {\rm sign}(\mu) \ ,  \ M_{\rm mes} \ , \ \Lambda \ , \
n_{\hat{q}} \ , \ n_{\hat{l}}
\eeq
In addition, one has to include as input the mass of the gravitino $\tilde{G}$
which, in this case is the lightest SUSY particle. This mass, $m_{\tilde
G}=F/(\sqrt{3} M_P)$ with $M_P$ the reduced Planck mass, will depend on an
additional free parameter $F$ which parameterizes the scale of the
full SUSY breaking and whose typical size is of 
${\cal O}(F_S)$  in direct 
mediation and much larger in secluded mediation. The choice
of this parameter, which plays a role only for the lifetime of the
next--to--lightest SUSY particle, is left to the user.  

\subsection*{2.6 Non--universality models}

mSUGRA, AMSB and GMSB are well defined models of which the possible
phenomenological consequences and experimental signatures have been widely
studied in the literature. However, in the absence of a truly fundamental
description of SUSY--breaking, none of these models should be
considered as {\sc the} compelling model. Unknown physics at the 
Planck and/or GUT scales renders some of the basic assumptions inherent to
the above scenarii rather uncertain: deviations from mSUGRA 
universality conditions at the high scale are generally expected in non-minimal
Supergravity \cite{weldon}, or in superstring \cite{ibanez} settings while
deviations from the initial AMSB scenario can be expected, due to extra 
anomaly contributions coming from the underlying fundamental theory \cite{AMSB1}
and deviations from minimal GMSB can occur due for instance to the breaking of
symmetries which protected direct coupling 
between messenger and MSSM matter fields. \s

To be on the safe side from the experimental point of view, it is therefore
wiser to allow for a departure from these models, and to study the
phenomenological implications of relaxing some defining assumptions.  
However, it is often
desirable to limit the number of extra free parameters, in order to retain a
reasonable amount of predictability when attempting detailed investigations of
possible signals of SUSY. Therefore, it is more interesting to relax only one
[or a few] assumption[s] at a time and study the phenomenological 
implications. Of
course, since there are many possible directions, this would lead to several
intermediate MSSMs between these constrained  models and the phenomenological
MSSM with 22 free parameters discussed in section 2.2.  \s

Taking the most studied mSUGRA as a reference model, 
we briefly discuss here some model cases where the universality conditions 
are naturally violated: \s

$i)$ non unification of the soft SUSY--breaking gaugino mass terms: 
\beq
M_1 (M_U) \neq M_2(M_U) \neq  M_3(M_U)
\eeq
This occurs for instance in Superstring motivated models in which the SUSY 
breaking is moduli dominated such as in the O--I and O--II models \cite{Pierre},
or in extra dimensional SUSY--GUT models in which the additional dimensions 
lead to the breaking of the large gauge symmetry and/or supersymmetry 
or in SUSY models where the breaking occurs through a non SU(5) singlet $F$
term; 
see Ref.~\cite{NM1} for phenomenology oriented discussions. \s

$ii)$ mSUGRA with non--unification of the two first and third generation scalar
masses: 
\beq
m_{0 \tilde{Q}}= m_{0 \tilde{L}} \cdots \neq m_{0 \tilde{q}} = m_{0 \tilde{l}} 
\cdots
\eeq
This occurs in models where the soft SUSY--breaking scalar masses at the 
GUT scale are influenced by the fermion Yukawa couplings. This is the case
for the so--called inverted mass hierarchy models \cite{NM2} where the 
scalar mass terms of the first two generations can be very heavy ${\cal O}
(10$ TeV),  while those of the third generation sfermions and the Higgs bosons
are rather light, solving thus the SUSY flavor and CP problems, which are 
related to the first two generations, while still satisfying naturalness 
constraints. \s 

$iii)$ mSUGRA-like models, but with non--universality of the sfermion and Higgs
boson scalar masses [keeping {\sl a priori} universal gaugino masses and
trilinear soft breaking couplings] . In a model-independent context, one 
can in principle consider any non-universality configuration which does not
 violate important constraints from FCNC, etc... Interestingly, such 
configurations occur naturally in models where the grand unification group has
 a rank strictly greater than $4$ \cite{manuel}. Starting from universal soft 
scalar masses, non-universality effects occur at intermediate scales, 
related to the spontaneous symmetry breaking down to the standard model 
gauge group, via D-term contributions corresponding to the extra U(1) 
symmetries \cite{KM}.      
For instance, in SO(10) SUSY GUT models with universal boundary conditions,
and assuming that the GUT group breaks directly to the SM group at the GUT
scale, one is lead to patterns of the form  \cite{KM, NM3}, 

\beq
m_{\tilde{Q}} =m_{\tilde{e}_R}= m_{\tilde{u}_R}  \neq m_{\tilde{d}_R} =
m_{\tilde{L}} \neq  M_{H_u} \neq M_{H_d}
\eeq

\noindent
where there are actually three free mass parameters, two of which can be taken
as the initial common soft masses for the sfermion and Higgs SO(10) multiplets,
the third being the D--term contribution  associated to the 
reduction of the group rank. \s  

In the present {\tt SuSpect} version such specific SUSY GUT models are not yet
available\footnote{These would require modified RGEs at least for some sectors,
inclusion of the effects of new heavy fields (right--handed (s)neutrinos for
instance), of threshold effects corresponding to the various scales of
successive symmetry breaking down to the SM, etc...}, but one can still have a
reasonable estimate of the non--universality effects by using the pMSSM option. 
Alternatively, one can also use the {\tt SuSpect} option which allows to
disconnect the Higgs sector from the sfermionic by introducing two additional
input parameters: the pseudoscalar Higgs boson mass $M_A$ and the higgsino mass
parameter $\mu$ [which have a  more direct ``physical" interpretation than the
scalar mass terms $M_{H_u}, M_{H_d}$].  This allows  to perform more general
phenomenological or experimental analyses; c.f. some LEP  and LHC Higgs
analyses \cite{Ex-nonuniv} or some recent Dark Matter studies
\cite{DM-nonuniv}. \s

$iv)$ Partially unified models where one relaxes one or a few parameters to fit
some collider zoo event or to analyze a phenomenological situation which
introduces new features. This is the case, for instance, for the light top
squark scenario which can be set by hand to discuss some theoretical issues
[such as baryogenesis in the MSSM \cite{baryo} for instance] or
phenomenological situations [such as new decay or production modes of top
squarks \cite{stop} for instance] .  \s 

An easy and practical way to implement these various non--unified or partially
unified scenarii, is to allow for the possibility of choosing all the soft 
SUSY--breaking parameters listed above for the phenomenological MSSM of section
2.2
[the 22 parameters except for $\tan \beta$] at the high--energy or GUT scale,
with the boundary conditions set by hand and chosen at will. One can even chose
the scale at which the boundary conditions are set to account for intermediate
scales. If this scale is the electroweak symmetry breaking scale, then we have
simply the MSSM with the soft SUSY--breaking parameters defined at the low
energy scale, i.e. the phenomenological MSSM.  All these options are
provided by our code.

\section*{3. The Particle Spectrum Calculation with Suspect} 

In this section, we discuss our procedure for calculating the SUSY and Higgs 
particle spectrum. We will take as example the sophisticated cases of the 
constrained MSSMs with universal boundary conditions at the high scale, mSUGRA
AMSB and GMSB, where all ingredients included in the {\tt SuSpect} algorithm 
are present: RGEs, radiative EWSB and calculation of the physical 
particle masses. We first describe the general algorithm,  then discuss 
the calculation of the soft SUSY--breaking terms, the determination of the
particle masses, the various  theoretical and phenomenological tests that we
impose on the model parameters. 

\subsection*{3.1 General algorithm}

As mentioned  previously, there are three main steps for the calculation of
the supersymmetric particle spectrum in constrained MSSMs, in addition to the
choice of the input parameters and the check of the particle spectrum: \s 

$i)$ Renormalization group evolution (RGE) of parameters [17--19], back and
forth between the low energy scales, such as $M_Z$ and the electroweak symmetry
breaking scale, and the high--energy scale, such as the GUT scale or the
messenger scale in GMSB models. This RGE is performed for  the SM gauge and
Yukawa couplings and for the soft SUSY--breaking terms (scalar and gaugino
masses, bilinear and trilinear couplings and $\tb$) and $\mu$. This procedure
has to be iterated in order to include SUSY threshold effects  or radiative
corrections due to Higgs and SUSY particles. \s 

$ii)$ The consistent implementation of (radiative) electroweak symmetry
breaking [20--23] and the calculation of $B$ and $|\mu|$ from the effective
scalar potential at one--loop level (plus the leading two--loop contributions).
Here, we use the tadpole method to include the loop corrections
\cite{potential}. The procedure has to be  iterated until a convergent value
for these two parameters is obtained.  [In the first step, the values of
$\mu^2$ and the electroweak symmetry  breaking scale are guessed by using the
tree--level  potential since no sparticle or Higgs mass has been calculated
yet.]\s 

$iii)$ Calculation of the pole masses of the Higgs bosons and the SUSY
particles, including the mixing between the current states and the radiative
corrections when they are important [24--31].  In this context, we will follow
mainly the content and notations of \cite{PBMZ}, to which we will refer as
PBMZ. The latter provides most of the necessary contributions at the one-loop
level for the Higgs and sparticle masses, while the leading two--loop
corrections to the (neutral) Higgs masses, calculated in 
Refs.~\cite{dsz,bdsz,bdsz2,dds}, are implemented in the form of a separate 
subroutine {\tt twoloophiggs.f} (provived by P. Slavich). \s 

\nn The general algorithm is depicted in Figure 1, and we will discuss the 
various steps in some detail in the following subsections. 

\begin{figure}[h!]
\vspace*{-.5cm}
\begin{picture}(1000,470)(10,0)

\Line(390,450)(450,450)
\ArrowLine(450,-60)(450,450)
\Line(450,-60)(390,-60)

\Line(10,460)(390,460)
\Line(10,390)(390,390)
\Line(10,460)(10,390)
\Line(390,460)(390,390)

\Text(200,450)[]{Low energy input:  $\alpha(M_Z),   
\alpha_S(M_Z)$, $M_t^{\rm pole}$, $M_{\tau}^{\rm pole}$,
$m_b^{\msbar}(m_b)$ ; $\tan \beta (M_Z)$}

\Text(200,430)[]{Radiative corrections $\Rightarrow$ $g_{1,2,3}^{\rm 
\overline{DR}}(M_Z)$, $Y_\tau^{\rm \overline{DR}} (M_Z), Y_b^{\rm 
\overline{DR}}(M_Z), Y_t^{\rm \overline{DR}} (M_Z)$}

\Text(200,410)[]{\it First iteration: no SUSY radiative corrections.} 

\ArrowLine(200,388)(200,372)

\Line(10,370)(390,370)
\Line(10,310)(390,310)
\Line(10,370)(10,310)
\Line(390,370)(390,310)

\Text(200,350)[]{One-- or two--loop 
RGE with choice: $\begin{array}{l}  g_1=g_2 \cdot \sqrt{3/5} \\ 
 M_{\rm GUT} \sim 2 \cdot 10^{16}~{\rm  GeV} \end{array}$}

\ArrowLine(200,308)(200,295)

\Text(200,285)[]{Choice of SUSY-breaking model (mSUGRA, GMSB, AMSB, or pMSSM).}
 
\Text(200,265)[]{Fix your high--energy input (mSUGRA: $m_0, m_{1/2}, A_0$, 
sign($\mu)$, etc...).} 

\ArrowLine(200,253)(200,242)

\Line(10,240)(390,240)
\Line(10,180)(390,180)
\Line(10,240)(10,180)
\Line(390,240)(390,180)

\Text(200,220)[]{Run down all parameters with RGE to $m_Z$ and 
$M_{\rm EWSB}$ scales} 
\Text(200,195)[]{\it First iteration: guess for $M_{\rm EWSB}$.}

\ArrowLine(200,178)(200,162)

\Line(10,160)(390,160)
\Line(10,100)(390,100)
\Line(10,160)(10,100)
\Line(390,160)(390,100)

\Text(200,150)[]{EWSB: $\mu^2, \mu B = F_{\rm non-linear}(m_{H_u}, m_{H_d}, 
\tan\beta, V_{\rm loop} )$}

\Text(200,130)[]{$V_{\rm loop} \equiv $ Effective potential at 1--loop with 
all masses.}

\Text(200,110)[]{{\it First iteration: $V_{loop}$ not included} }

\Line(390,130)(420,130)
\ArrowLine(420,30)(420,130)
\Line(420,30)(390,30)

\ArrowLine(200,98)(200,90)

\Text(200,85)[]{Check of consistent EWSB ($\mu$ convergence, no tachyons, 
simple CCB/UFB, etc...) } 

\ArrowLine(200,75)(200,62)

\Line(10,60)(390,60)
\Line(10,0)(390,0)
\Line(10,60)(10,0)
\Line(390,60)(390,0)

\Text(200,50)[]{Diagonalization of mass matrices and calculation of masses /  
couplings}

\Text(200,30)[]{Radiative corrections to the physical Higgs, sfermions, gaugino
masses.}

\Text(200,10)[]{\it First iteration: no radiative corrections.} 

\ArrowLine(200,-2)(200,-17)

\Line(10,-20)(390,-20)
\Line(10,-85)(390,-85)
\Line(10,-20)(10,-85)
\Line(390,-20)(390,-85)

\Text(200,-30)[]{Check of a reasonable spectrum:} 

\Text(200,-45)[]{-- no tachyonic masses (from RGE, EWSB or mix), 
\ \hfill } 

\Text(200,-60)[]{--information provided on fine-tuning, CCB/UFB 
conditions,\hfill }

\Text(240,-75)[]{--calculation of MSSM contributions to: $\Delta\rho$, $(g-2)$, 
$b \to s\gamma$. \ \ \ \ \ \ \ \ \ \ \ \ \ \ \ \ \ \ \ \ \hfill}
\end{picture}

\vspace*{3.2cm}

\noindent {\it Figure 1: Iterative algorithm for the calculation of the SUSY
particle spectrum in {\tt SuSpect} from the choice of input (first step) to the
check of the  spectrum (last step). The steps are detailed in the various 
subsections. The EWSB iteration [calculationally fast] on $\mu$ is performed
until  $|\mu_{i}-\mu_{i-1}| \leq \epsilon |\mu_i|$ (with $\epsilon \sim 10^{-4}$)
while  the RG/RC ``long" iteration [calculationally longer] is performed
until some (user specified) accuracy is reached}
\vspace*{-.2cm}
\end{figure}

\subsection*{3.2 Calculation of the MSSM parameters at the low scale}

Prior to the calculation of all the relevant terms entering the physical mass
spectrum calculation at a low (EWSB) scale, such as typically the soft
SUSY--breaking terms in constrained models with boundary conditions at the
unification scale, the first important stage is to define a choice of
low--energy input parameters and how they are extracted from the present
experimental data. This is the purpose of the next subsection.  

\subsubsection*{3.2.1 Choice and treatment of the SM inputs}

 The low--energy (weak) scale boundary conditions set the gauge and Yukawa
couplings by matching the running MSSM parameters to the experimental data at a
chosen renormalization scale, usually $Q=\polemz$. This step is quite involved,
requiring to subtract the radiative corrections from the experimental data in
order to arrive at the $\drbar$--renormalized MSSM parameters. Using the
formulae of mainly Ref.~\cite{PBMZ} and references therein, the MSSM $\drbar$
gauge couplings $g_1 \,, g_2 \,, g_3$ and the electroweak parameter $v$ can be
computed at $Q=\polemz$ from a set of four experimental input parameters. \s

The latter can be chosen as: $G_F$, the Fermi constant determined from the muon
decay; $\polemz$, the pole mass of the $Z$ boson; $\alpha_{\rm
em}(\polemz)^{\msbar}$, the five--flavour SM electromagnetic coupling at the
scale $\polemz$ in the $\msbar$ scheme; $\as(\polemz)^{\msbar}$, the
five--flavour SM strong coupling at the scale $\polemz$ in the $\msbar$ scheme.
Then, the running couplings $g_2 \,, g_1$ are connected to the running
$Z$--boson mass $\mz$ by the relation:
\be
\label{mzrun}
\mz^2 = \polemz^2 + {\rm Re}\, \Pi_{ZZ}^T(\polemz^2) 
= \frac{1}{4}\,(g^2_1 + g^2_2)\,v^2,
\ee
where $\Pi_{ZZ}^T(\polemz^2)$ is the transverse part of the $Z$ boson
self--energy computed at a squared external momentum equal to the
squared pole $Z$ boson mass. 
[Note that according to this input choice, the $W$ boson mass 
and the parameter $\sin^2\theta_W$ are 
not free parameters but can be derived when needed, e.g. in the
$\drbar$ scheme, 
from the above defined input parameters using appropriate
relations including radiative corrections]. \s

The next MSSM input parameter is $\tb \equiv v_u/v_d$, the ratio of the  vacuum
expectation values (VEVs) of the two MSSM neutral Higgs fields, $H_d^0$ and
$H_u^0$, defined also at the scale $Q=M_Z$, with $v_d^2 + v_u^2 \equiv v^2$.\s 

Then the Yukawa couplings $Y_u$ ($u = u,c,t$) for the up--type quarks,
$Y_d$ ($d = d,s,b$) for the down--type quarks and $Y_{\ell}$ ($\ell = e,\mu,
\tau$) for the leptons are determined 
in the $\drbar$ scheme from the corresponding
running fermion masses as
\be
\label{yukawas}
Y_u = \frac{\sq2\,m_u}{v\,\sin\beta}\,,\;\;\;\;\;\;
Y_d = \frac{\sq2\,m_d}{v\,\cos\beta}\,,\;\;\;\;\;\;
Y_{\ell} = \frac{\sq2\,m_{\ell}}{v\,\cos\beta}\,,
\ee
The $\drbar$ running fermion masses (see e.g. 
Refs.~\cite{qqcd,broad,runmass,mbdr}) 
$m_f$ (with $f=u,d,\ell$) in eq.~(\ref{yukawas}) can be derived at the 
one--loop level from the corresponding pole masses $M_f$ through the relation
\be
m_f = M_f + \Sigma_f(M_f)
\label{mfMf}
\ee
where $\Sigma_f(M_f)$ is the one--loop fermion self--energy computed
at an external momentum equal to the pole mass. In the case of the top
quark, the self--energy includes also the leading two--loop standard
QCD corrections \cite{broad}
\be
m_t = \polemt + \Sigma_t(\polemt) + (\Delta m_t)^{\rm 2-loop,\,QCD}
\label{corrmt}
\ee
where the precise expression for $(\Delta m_t)^{\rm 2-loop,\,QCD}$
depends on the renormalization scheme in which the parameters entering
the one--loop self--energy $\Sigma_t$ are expressed. Concerning
the bottom quark, we follow now the SUSY {\em Les Houches Accord} 
\cite{slha} which prescribes to take as input the SM running
mass in the $\msbar$ scheme, $m_b(m_b)^{\msbar}$. In addition, a
``resummation'' procedure is required (see e.g. Ref.~\cite{resum}) in
order to properly take into account the large QCD corrections, as well
as the $\tb$--enhanced SUSY corrections \cite{hrs}, to the relation
between the input bottom mass and the corresponding MSSM, $\drbar$
Yukawa coupling. We extract the latter, via eq.~(\ref{yukawas}), from
the MSSM, $\drbar$ bottom mass $\widehat{m}_b$, defined at the scale
$Q=\polemz$ by the following matching condition:
\be
\label{mbmssm}
\widehat{m}_b \equiv  m_b(\polemz)^{\drbar}_{\rm MSSM} =
\frac{\ov{m}_b}{1-\Delta_b}
\label{mbresum}
\ee
where $\ov{m}_b \equiv m_b(\polemz)^{\drbar}_{\rm SM}$ is the SM,
$\drbar$ bottom mass, obtained by evolving $m_b(m_b)^{\msbar}$ up to
the scale $Q=\polemz$ with the appropriate RGE, in order to resum the
QCD corrections, and then converting it to the $\drbar$ scheme;
$\Delta_b \equiv \Sigma_b(\widehat{m}_b)/\widehat{m}_b$ accounts for
the remaining non--gluonic corrections, some of which are enhanced by
a factor $\tb$. It has been shown \cite{resum} that defining the
running MSSM bottom mass as in eq.~(\ref{mbmssm}) guarantees that the
large threshold corrections of ${\cal O}(\as \tb)^n$ are included in
$\widehat{m}_b$ to all orders in the perturbative expansion. Concerning 
the $\tau$ lepton, the only $\tb$--enhanced corrections to be
included are those controlled by the electroweak gauge couplings,
from chargino--slepton loops.\s

Concretely the (present) reference values of the
SM input parameters at the weak scale 
are fixed as follow. 
We take for the electroweak and strong
parameters \cite{PDG}:
\[
G_F = 1.16639\, 10^{-5}\;{\rm GeV}^{-2},\;\;\;\;\;
\polemz = 91.1876\; {\rm GeV},
\]
\be
\label{inputew}
\alpha_{\rm em}^{-1}(\polemz)^{\msbar} = 127.934 ,\;\;\;\;\;
\as(\polemz)^{\msbar} = 0.1172, \
\ee
and for the third--generation fermion masses the values \cite{PDG,toptev}:
\be
\label{defmfi}
\polemt = 178.0 \; {\rm GeV},\ \ \
m_b(m_b)^{\msbar} = 4.25\; {\rm GeV},\ \ \
M_{\tau} = 1.777 \; {\rm GeV}.
\ee
Of course, for maximal flexibility, all those SM input default values
in eqs.~(\ref{inputew},\ref{defmfi}) can be changed at will 
in the input file. \s

Once the Supersymmetric particle spectrum has been obtained [see below], we
include all the important SUSY radiative corrections to the gauge and Yukawa
couplings, via their above defining relations from input parameters, e.g.
eqs.~(\ref{mzrun},\ref{corrmt}) or (\ref{mbresum}), where
the SUSY particle masses enter the expression of e.g. $\Pi_{ZZ}^T$
in eq.~(\ref{mzrun}) and $\Sigma_t$ in eq.~(\ref{corrmt}). 
This necessarily implies an iterative procedure, since the
values of the SUSY particle spectrum depend, among
other thing, on the precise 
values of the gauge and Yukawa couplings via the RGE typically.
The iteration is to be performed until a sufficiently
stable final SUSY spectrum is obtained [and where the very first iteration
is done with only SM radiative corrections included in 
eqs.~(\ref{mzrun},\ref{corrmt}), before SUSY particle masses are
defined, see the general algorithm overview in Fig.~1]. 
  In the case of the Yukawa couplings, we include all relevant
SUSY corrections to the third generation fermion masses. For the bottom quark
($\tau$ lepton) mass,  we include the SUSY--QCD and stop--chargino
(sneutrino--chargino) one--loop corrections at zero--momentum transfer
\cite{resum}
which, according to PBMZ, is an extremely good approximation. 
These corrections to the
$b$ and $\tau$ masses are enhanced by terms $\propto \mu \tb$ and can be rather
large. 
For the top quark, the inclusion of only the leading corrections at zero
momentum transfer is not an accurate approximation, and we include the 
full one--loop SUSY--QCD [i.e stop and gluino loops] and electroweak [i.e. 
with gauge, Higgs boson and  chargino/neutralino exchange] corrections 
{\it \`a la} PBMZ \cite{PBMZ}.

\subsubsection*{3.2.2 Renormalization Group Evolution}

All gauge and (third generation) Yukawa couplings are then evolved up to the
GUT scale using the two--loop MSSM RGEs \cite{RGE2,drbar} in the $\drbar$
scheme, with the contribution of all the MSSM particles
in the relevant beta--functions. The GUT scale, $M_{\rm GUT} \simeq 2 \cdot
10^{16}$ GeV
can be either fixed by hand or, by appropriate user's choice in the input
file,  calculated consistently to be the scale at which the electroweak gauge
coupling constants [with the adequate normalization] unify, $g_1 = g_2 \cdot
\sqrt{3/5}$. In contrast, we do not enforce exact $g_2 = g_3$  unification at
the GUT scale and assume that the small discrepancy, of at most a few percent,
is accounted for by unknown GUT--scale threshold corrections
\cite{gutthresh}.\s 

One can then chose the parameter $\tan\beta$, given at the scale $M_Z$, the
sign of the $\mu$ parameter and, depending on the chosen model, the high energy
and the low energy input. For instance, one can set the high--energy scale
$E_{\rm High}$, which in mSUGRA or AMSB can be either forced to be $M_{\rm
GUT}$ [the scale at which $g_1$ and $g_2$ unify] or chosen at will [any
particular intermediate scale between $M_Z$ and $M_{\rm GUT}$ can be allowed
in general  and in the case of the GMSB model this scale corresponds to the
messenger  scale $M_{\rm mes}$]. Similarly the low energy scale $E_{\rm Low}$,
where the RGEs  start or end may be chosen [it is in general taken to be
$M_Z$ or the EWSB scale to be discussed later]. The additional input in the
various models are: 
\begin{itemize}
\vspace*{-2mm}
\item \underline{mSUGRA}: the universal trilinear coupling $A_0$, the common 
scalar mass $m_0$ and the common gaugino mass $m_{1/2}$, all defined at the 
scale $M_{\rm GUT}$.  
\vspace*{-2mm}

\item \underline{AMSB}: the common scalar mass $m_0$, the gravitino mass 
$m_{3/2}$ and  the set of coefficients $c_{S_i}$ for the non--anomalous 
contributions, to be as general as possible. 
\vspace*{-2mm}

\item \underline{GMSB}: the scale  $\Lambda$, the messenger scale $M_{\rm mes}$ 
which corresponds to $E_{\rm High}$, as well as the numbers of messengers $n_q$ 
and $n_l$. 
\vspace*{-2mm}

\item \underline{pMSSM} with boundary conditions: the 
various soft SUSY--breaking parameters listed in section 2.2 [21 parameters in 
total, in addition to $\tan\beta$] defined at the scale $E_{\rm High}$. These
input can also be chosen at will at the low--energy scale $E_{\rm Low}$ 
which has also to be provided as a necessary input. For this pMSSM input 
choice we stress that all the (scale--dependent) soft SUSY--breaking
parameters are implicitly understood to be defined at the given
arbitrary low scale  $E_{\rm Low}$. In particular, the code will
perform RGE consistently between the scale $Q=M_Z$ and  $Q=E_{\rm Low}$,
to match e.g. with the SM input defined at $Q=m_Z$, as well
as the RGE between  $Q=E_{\rm Low}$ and the EWSB scale $M_{\rm EWSB}$,
while
the RGE from a high scale $E_{\rm High}$ down to $M_{\rm EWSB}$ 
are not relevant in this case and    
are thus switched off. Note also that here, 
a very convenient option is provided
which allows to trade the input parameters $M_{H_u}^2$ and  $M_{H_d}^2$ with
the more ``physical" parameters $M_A$ (pole mass) and 
$\mu(E_{\rm low})$ [again in such a way that  EWSB
is consistently realized, with a warning flag whenever it is not the case].
\vspace*{-2mm}
\end{itemize}

Given these boundary conditions, all the soft SUSY breaking parameters and 
couplings are evolved down to the EWSB and $M_Z$ scales, 
using either the one-- or (preferably) 
full two--loop RGE options\footnote{The full 
two--loop RGEs in the MSSM 
\cite{Martin}, have been implemented for the slepton and quark masses
in the latest version of \suspect\ . In the previous versions, 
the two--loop RGEs were included for all terms except for the latter
soft SUSY--breaking masses. These higher order effects
can have a non-negligible impact \cite{compsabi} in some
regions of the MSSM parameter space.}. 
Our
default  choice for the EWSB scale is the geometric mean of the two top 
squark {\em running} 
masses (in the $\drbar$ scheme),  \beq
M_{\rm EWSB} = (m_{\tilde{t}_1} m_{\tilde{t}_2})^ {1/2}
\eeq
which minimizes the scale dependence of the one--loop effective
potential  \cite{Vscale} discussed below [at first iteration
where the stop masses have not yet been
calculated, we use the geometric mean of the soft SUSY--breaking stop masses
instead as a first guess]. Note, however, that any other arbitrary values of
the EWSB scale can be chosen easily by an appropriate input setting.
Since $\tb$ is defined at $M_Z$, the vevs have to be
evolved down  from $M_{\rm EWSB}$ to $M_Z$. \s

Once the SUSY spectrum is calculated [see below], the heavy
(s)particles are taken to contribute to reevaluate the gauge and Yukawa
couplings 
at the scale $Q=M_Z$, as discussed in the previous subsection, and
the necessary iterative procedure thus also includes
the RGEs. \s

\subsubsection*{3.2.3 Electroweak Symmetry Breaking} 

At some stage, we require that the electroweak symmetry is broken radiatively
and use eq.~(\ref{eq:ewsb}) to determine the parameters $\mu^2(M_{\rm EWSB})$
and $B(M_{\rm EWSB})$. It is well known that the one--loop radiative
corrections to the Higgs potential play a major role in determining the values
of these two parameters, which at tree level are given  in terms of the soft 
SUSY--breaking masses of the two Higgs doublet fields. We treat these
corrections using the tadpole method. This means that we can still use
eq.~(\ref{eq:ewsb}) to determine $\mu^2(M_{\rm EWSB})$, one simply has to add
one--loop tadpole corrections $t_u$, $t_d$ \cite{potential,PBMZ} 
\beq
m^2_{H_u} \to m^2_{H_u} -t_u/v_u \ {\rm and} \  m^2_{H_d} \to m^2_{H_d}
-t_d/v_d \eeq
We include the dominant third generation fermion/sfermion loops, as well as 
sub--dominant contributions from sfermions of the
first two generations, gauge bosons, the MSSM Higgs bosons, charginos and
neutralinos\footnote{The contributions of the charginos and neutralinos can be
rather sizable and are very important to minimize the scale dependence of the
one--loop effective potential \cite{Spanos}.}, with the running parameters
evaluated at $M_{\rm EWSB}$. Note that we also include the leading
two--loop tadpole corrections, as calculated in Refs. \cite{dsz,bdsz,bdsz2},
to be consistent in particular with the subsequent calculation of
the physical neutral Higgs masses including the leading two--loop
contributions. \s

As far as the determination of $\mu^2$ and $B\mu$ is concerned, this is
equivalent (at the considered loop level) to computing 
the full effective potential at scale $M_{\rm
EWSB}$. Since $|\mu|$ and $B$ affect the masses of some (s)particles appearing
in these corrections, this gives a non--linear equation for $|\mu|$ (see  
Fig.~1), which is  solved by a standard iteration algorithm until stability is
reached and a consistent value of $\mu$ is obtained. From a practical point of
view, this requires only three or four iterations for an accuracy of ${\cal
O}(10^{-4})$, if one starts from the values of $|\mu|$ and $B$ as determined
from minimization of the RG--improved tree--level potential at scale $M_{\rm
EWSB}$ and the procedure is extremely fast in CPU as compared to the (iterated)
RGE calculation.\s

At this stage, \sus\ includes a  check on whether the complete scalar potential
has charge and/or  color breaking (CCB) minima  which can be lower than the
electroweak minimum, or whether the tree--level scalar potential is unbounded
from below (UFB). In the present version of the code, we consider only the
following simple  (tree--level) criteria \cite{CCBold}
\beq
{\rm CCB1}: && A^2_f < 3 \, (m^2_{\tilde f_L} +m^2_{\tilde f_R}+ \mu^2 +
m^2_{H_u} ). \label{CCBcons} \\
{\rm UFB1}: && m^2_{H_u} + m^2_{H_d} \geq 2 |B\mu| \ \ {\rm at\, scale} \  
Q^2 >M_{\rm EWSB}^2  \label{UFBcons}
\eeq
where $f$ denotes any of the three fermion generations.  Eq.~(\ref{CCBcons})
ensures that there is no deep CCB breaking minimum [due to very small Yukawa
couplings]  in some D--flat directions. One can either take this as a
consistency necessary constraint on the MSSM parameters, or disregard it
appealing  to the fact that such minima are usually well separated from the
electroweak minimum so that the latter can be reasonably stable at
cosmological  scales\footnote{But one would still lack for a compelling reason
why the EW minimum is chosen in the first place.}. For the third generation and
in particular in the top sector, the CCB minimum is not much deeper than the
electroweak minimum, since $Y_t$ is not very small,  and not much separated
from it. In this case one should  apply eq.~(\ref{CCBcons}) with some caution
since tunneling effects can be important. On the other hand the
``boundedness--from--below" condition of eq.~(\ref{UFBcons})  is actually an
indication of possible dangerous non physical minima which could form when
radiative corrections are included. At any rate, since both
eqs.~(\ref{CCBcons}) and (\ref{UFBcons}) are merely tree--level conditions,
they should be checked at the highest energy scale.  Note that 
all subsequent calculations (i.e. the sparticle and Higgs masses)
are  still performed, even if these
conditions are not fulfilled, but a warning  flag is given in the output file. 
An upcoming version of \sus\ will have
more sophisticated treatments, taking into account loop corrections \cite{CCB}
as well as the geometric configurations of the true minima
\cite{lemouel,Kusenko} as will be discussed later.  \s 

Finally, \sus\ provides appropriate (rejection) flags
for any input choice in the parameter space which lead to 
tachyonic pseudo--scalar Higgs boson or sfermion masses:
\beq
{\rm No\,  Tachyon}: && M_A^2 >0 \ \ , \ \ m^2_{\tilde f} >0 . 
\eeq 
The electroweak 
symmetry breaking  mechanism is assumed to be consistent when all these 
conditions are satisfied. Note however that for some input parameter
choice such problems may occur
temporarily i.e. before a sufficiently stable SUSY spectrum is obtained,
and accordingly the \sus\ algorithm allows the calculation to be 
nevertheless performed until the very last iteration before eventually
issuing a rejection flag. 

\subsection*{3.3 Calculation of the physical particle masses} 

Once all the soft SUSY--breaking terms are obtained and eventually EWSB is
radiatively realized [as should be the case in unified models] one can then
calculate all the physical particle masses. 
As already mentioned the  whole procedure (namely, RGE +
EWSB + spectrum calculation) is iterated a number of time until stability is
reached (see the overall algorithm in Fig 1), in order to take into account
realistic and stable particle masses in the ``threshold" corrections 
using the expressions given in Ref.~\cite{PBMZ}, 
to the gauge and Yukawa couplings at the scale $Q=M_Z$.\s

Our conventions for the mass matrices in the gaugino, sfermion and
Higgs sectors will be specified below. We basically follow the conventions of
PBMZ with some important exceptions: $(i)$ The $\mu$ parameter is defined with
the opposite sign (see below). $(ii)$ The vevs are different by a factor
$\sqrt{2}$ and  in our case $v=174.1$ GeV. $(iii)$ The sfermion masses are
defined such that $\tilde{f}_1$ and $\tilde{f}_2$ are, respectively, the
lightest and the heaviest one. $(iv)$ The matrices diagonalizing the
chargino and neutralino mass matrices are taken to be real.\s

For the calculation of the physical masses and the implementation of the 
radiative corrections, the various sectors of the MSSM are then treated as 
follows [with some details on the notation and conventions we use]:

\subsubsection*{3.3.1. The sfermion sector}

In the third generation sfermion sector [$\tilde{t},\tilde{b}, \tilde{\tau} $],
mixing between ``left'' and ``right'' current eigenstates is included 
\cite{sfmix}. The
radiatively corrected running $\drbar$ 
fermion masses [essentially the Yukawa coupling
times vevs] at scale $M_{\rm EWSB}$ are employed in the sfermion mass matrices
[this is important at large $\tb$, where these corrections can be quite
sizable].  As mentioned above, contrary to PBMZ, the masses are defined such
that $m_{\tilde{f}_1}$ and $m_{\tilde{f}_2}$ correspond to the mass of
respectively, the lightest and the heaviest sfermion and therefore care
should be made in interpreting the sfermion mixing angle $\theta_{\tilde f}$ as
compared to PBMZ.  [Note that a protection which prevents negative
mass squared for third generation sfermions in the presence of too large
mixing is provided.] The sfermion mass matrices 
are given  by: 
\begin{eqnarray}
 M_{\tilde{f}}^2 \ = \ 
\left[ \begin{array}{cc} m_{\tilde{f}_L}^2 + ( I^3_f - e_{f} s_W^2) 
M_Z^2 \cos2\beta + m_f^2 & m_f (A_f - \mu r_f) 
\\ m_f (A_f - \mu r_f)  & m_{\tilde{f}_R}^2 - e_{f} s_W^2
M_Z^2 \cos2\beta + m_f^2 \end{array} \right]
\label{defMsf}
\end{eqnarray}
where $m_{\tilde{f}_{L,R}}, A_f, \mu$ and $m_f$ are respectively, 
the $\overline{\rm DR}$ soft SUSY scalar masses, trilinear couplings, higgsino
mass parameter and running fermion masses at the scale $M_{\rm EWSB}$ 
and $r_{b} = r_\tau =1/r_t= \tb$. These matrices are diagonalized by orthogonal
matrices; the mixing angles $\theta_f$ and the squark eigenstate masses are
determined uniquely, at the relevant energy scales\footnote{Note that the 
sfermion masses and mixing angles are also calculated for consistency by \sus\ 
at the $M_Z$ scale as well since they enter the various radiative corrections 
to the gauge and Yukawa couplings.} by
\begin{eqnarray}
\tan 2\theta_f = \frac{2 m_f (A_f -\mu r_f)}{m_{\tilde{f}_L}^2 -m_{\tilde{f}_R}^2
+ I^3_f M_Z^2 \cos2\beta}
\end{eqnarray}
or
\begin{eqnarray}
\tan \theta_f =  \frac{m_{\tilde{f}_1}^2 - m_{\tilde{f}_2}^2 + 
m_{\tilde{f}_R}^2 -m_{\tilde{f}_L}^2 - I^3_f M_Z^2 \cos2\beta }{2 m_f (A_f -\mu
r_f)} 
\end{eqnarray}
and 
\begin{eqnarray}
m_{\tilde{f}_{1,2}}^2 = m_f^2 + \frac{1}{2} \left[ 
m_{ \tilde{f}_L}^2 + m_{\tilde{f}_R}^2 \mp \sqrt{
(m_{\tilde{f}_L}^2 - m_{\tilde{f}_R}^2 + I^3_f M_Z^2 \cos2\beta)^2 + 4m_f^2 (A_f
-\mu r_f)^2 } 
\right] 
\end{eqnarray}
Our convention for the mixing angles is
\begin{eqnarray}
\left(\begin{array}{c} \tilde{f}_1 \\ \tilde{f}_2 \end{array} \right) &=&
 \left[ \begin{array}{cc} \cos \theta_f & \sin \theta_f \\
                           - \sin \theta_f & \cos \theta_f  \end{array} \right]
\left(\begin{array}{c} \tilde{f}_L \\ \tilde{f}_R \end{array} \right)
\end{eqnarray}
where $\tilde{f}_i$ denote the mass eigenstates, and $\tilde{f}_{L,R}$ the
chiral states. Equations (45, 46)  ensure simultaneously that 
each $m_{\tilde{f}_i}$ given in (44) is indeed the mass corresponding to
the state $\tilde{f}_i$ and that $\tilde{f}_1$ is always the lightest state. In
practice,  the possible 
ambiguity, when $\theta_f$ is determined from 
$\frac12 \arctan (\tan 2 \theta_f)$, is lifted in \sus\ by requiring that 
$\tan \theta_f$ has always an opposite sign to $A_f -\mu r_f$,  as can be seen
from eq.~(45)\footnote{In other parts of the code where sfermion masses and 
mixing angles are used internally, for instance when
entering in loop corrections, only the first equation in (44) is used to
determine the mixing angles. In this case the consistency is guaranteed through
the calculation of the eigenmasses directly from the mixing angles 
in the form $(M_{\tilde{f}}^2)_{11} \cos^2 \theta_f +  (M_{\tilde{f}}^2)_{22}
\sin^2 \theta_f \pm (M_{\tilde{f}}^2)_{12} \sin 2 \theta_f$, rather than from
(45). Of course, the convention 
$m_{\tilde{f}_{1}}^2 < m_{\tilde{f}_{2}}^2$ is not necessarily valid here.}.\s

The radiative corrections to the sfermion masses are included according to
Ref.~\cite{PBMZ}, i.e. only the QCD corrections for the superpartners of
light quarks [including the bottom squark] plus the leading electroweak
corrections to the two top squarks; the small electroweak radiative corrections
to the slepton masses [which according to PBMZ are at the level of one percent]
have been neglected in the present version. 

\subsubsection*{3.3.2 The gaugino sector}

The $2 \times 2$ chargino and $4\times 4$ neutralino mass matrices  
depend on the $\overline{\rm DR}$ parameters $M_1, M_2, \mu$ at the scale
$M_{\rm EWSB}$ and on $\tb$. The chargino mass matrix given by: 
\begin{eqnarray}
{\cal M}_C = \left[ \begin{array}{cc} M_2 & \sqrt{2}M_W \sin \beta
\\ \sqrt{2}M_W \cos \beta & \mu \end{array} \right]
\label{defMc}
\end{eqnarray}
is diagonalized by two real matrices $U$ and $V$. The chargino masses are
obtained analytically, with the convention that $\chi_1^+$ is the lightest
state. \s

The neutralino mass matrix, in the $(-i\tilde{B},
-i\tilde{W}_3, \tilde{H}^0_1,$ $\tilde{H}^0_2)$ basis, has the form  
\begin{eqnarray}
{\cal M}_N = \left[ \begin{array}{cccc}
M_1 & 0 & -M_Z s_W \cos\beta & M_Z  s_W \sin\beta \\
0   & M_2 & M_Z c_W \cos\beta & -M_Z  c_W \sin\beta \\
-M_Z s_W \cos\beta & M_Z  c_W \cos\beta & 0 & -\mu \\
M_Z s_W \sin \beta & -M_Z  c_W \sin\beta & -\mu & 0
\end{array} \right]
\label{defMn}
\end{eqnarray}
It is diagonalized using analytical formulae \cite{egypte} by a single matrix
$Z$ which is chosen to be real, leading to the fact that some (in general one)
of the neutralino eigenvalues is negative. The physical masses
are then the absolute values of these eigenvalues with some reordering such
that the neutralinos $\chi_{1,2,3,4}^0$ are heavier with increasing subscript
and $\chi_1^0$ is the lightest neutralino. \s

For the gluino, the  running $\overline{\rm DR}$ mass $m_{\tilde g}$ at scale
$M_{\rm EWSB}$ is identified with $M_3(M_{\rm EWSB})$
\beq
m_{\tilde g}^{\rm tree}= M_3(M_{\rm EWSB})
\eeq
The full one--loop QCD radiative corrections to the gluino mass are
incorporated \cite{drbar}, while in the charginos/neutralinos case the
radiative corrections to the masses are simply included in the
gaugino and higgsino limits, which is a very good approximation \cite{PBMZ}.

\subsubsection*{3.3.3. The Higgs sector}

In the latest version of \sus\ we have performed an up--to--date and more 
precise determination of the physical Higgs masses including, in particular, 
the leading two--loop contributions \cite{dsz,bdsz,bdsz2,dds}
to the neutral Higgs masses and tadpoles. 
The running $\overline{\rm DR}$ mass of the pseudoscalar Higgs boson at the
scale $M_{\rm EWSB}$, $\bar{M}_A$, is obtained from the soft SUSY--breaking
Higgs mass terms evolved from RGE at the scale $M_{\rm EWSB}$ and including the
loop tadpole corrections \cite{PBMZ}
\beq
\bar{M}_A^2 (M_{\rm EWSB}) &=& \frac{1}{\cos 2\beta} \bigg( m_{H_d}^2-
\frac{t_d}{v_d} - 
m_{H_u}^2 + \frac{t_u}{v_u} \bigg)-\bar{M}_Z^2 + \sin^2\beta \frac{t_d}{v_d} + 
\cos^2 \beta \, \frac{t_u}{v_u}  
\eeq
This mass, together with the $Z$ boson mass $\bar{M}_Z$
at scale $M_{\rm EWSB}$, are then used as inputs in the CP--even Higgs boson 
$2 \times 2$ mass matrix ${\cal M}_S$. Including the dominant contributions of 
the self--energies of the unrotated CP--even neutral Higgs fields $H_u^0$ and 
$H_d^0$ (as well as the tadpole contributions), this matrix reads at a given 
scale $q^2$ 
\begin{eqnarray}
{\cal M}^S(q^2) = \left[ 
\begin{array}{cccc}
\bar{M}_Z^2\cos\beta^2 + \bar{M}_A^2 \sin^2\beta - s_{11} (q^2) &
- \frac{1}{2}(\bar{M}_Z^2 + \bar{M}_A^2) \sin2\beta - s_{12}(q^2) \\
- \frac{1}{2} (\bar{M}_Z^2+ \bar{M}_A^2) \sin2\beta - s_{12} (q^2) 
& \bar{M}_Z^2 \sin^2\beta + \bar{M}_A^2 \cos^2\beta - s_{22}(q^2) 
\end{array} \right]
\end{eqnarray}
One obtains the running CP--even Higgs boson masses in terms of the matrix 
elements ${\cal M}^S_{ij}$
\beq
\bar{M}_{h,H}^2 = \frac{1}{2} \bigg[ {\cal M}^S_{11}+{\cal M}^S_{22} \mp
\sqrt{ ( {\cal M}^S_{11}-{\cal M}^S_{22})^2+ 4 ({\cal M}^S_{12})^2 } \bigg] 
\eeq
The mixing angle $\alpha$ which diagonalizes the matrix ${\cal M}^S$ and 
rotates 
the fields $H^0_u, H_d^0$ into the physical CP--even Higgs boson fields $h, H$ 
\begin{eqnarray} 
 \left( \begin{array}{c} H \\ h \end{array}  \right)  = 
 \left( \begin{array}{cc} \cos \alpha & \sin \alpha  \\ -\sin \alpha & 
\cos\alpha \end{array} \right) \ \left( \begin{array}{c} H^0_d \\ H_u^0  
\end{array} \right) 
\end{eqnarray}
is given by 
\beq
\sin 2 \alpha = \frac{ 2 {\cal M}^S_{12} }{ \bar{M}_H^2- \bar{M}_h^2 } \ \ , \ 
\cos 2 \alpha = \frac{ {\cal M}^S_{11} - {\cal M}^S_{22} }{ \bar{M}_H^2-
\bar{M}_h^2 }
\ \ \ \bigg( -\frac{\pi}{2} < \alpha < \frac{\pi}{2} \bigg) 
\eeq
The running charged Higgs boson mass at the EWSB scale is given by
\beq
\bar{M}_{H^\pm}^2&=& \bar{M}_A^2 + \bar{M}_W^2 - \sin^2 \beta \, \frac{t_d}{v_d} - 
\cos^2 \beta \, \frac{t_u}{v_u} 
\eeq
The pole masses of all the Higgs bosons are then obtained by including the
self--energy corrections evaluated at the masses of the Higgs bosons 
themselves. \s

In the evaluation of the radiative corrections in the MSSM Higgs sector which 
are known to be very important \cite{RCH}, we have made several options
available: \s

$(i)$ Approximate one--loop and two--loop contributions to the self--energies
(and tadpole) corrections $s_{ij}$ in the mass matrix ${\cal M}^S$. These
expressions, given in Ref.~\cite{FeynHiggsFast}, provide a rather good
approximation (at the few percent level) for the masses $M_h$ and $M_H$ and the
angle $\alpha$ at least for a reasonable range and not too extreme values of
input parameters.  Since  it makes the program running faster, this
approximation may be convenient in cases where a very precise determination of
the Higgs masses may not be mandatory.\s

$(ii)$ A full one--loop calculation of the Higgs masses, following essentially
the expressions of Ref.~\cite{PBMZ}. In this case, the
tadpoles are consistently evaluated at the one--loop level only.\s

$(iii)$ A full one--loop plus 
leading two-loop calculation controlled by the third
generation Yukawa couplings and the strong gauge coupling, derived
in Refs.~\cite{dsz,bdsz,bdsz2,dds}. 
Note that the calculation is entirely performed
in the $\drbar$ scheme, therefore
different from the on-shell scheme calculation
performed in  {\tt FeynHiggsFast} of Heinemeyer, Hollik and Weiglein 
\cite{FHF,FeynHiggsFast}.  The comparison of results from $(ii)$ and
$(iii)$ can be useful in order to see the effects from two--loop
contributions. For a recent detailed analysis of the Higgs masses
in various models and a comparison between \sus\, and other codes,
we refer to Ref.~\cite{adkps}.

\subsection*{3.4 Theoretical and experimental constraints on the spectra}

Once the SUSY and Higgs spectrum is calculated, one can check that some
theoretical and experimental requirements are fulfilled. Examples of theory
requirements are for instance, the absence of charge and color breaking (CCB)
minima and that the potential is not unbounded from below (UFB), the absence of
too much fine--tuning (FT) in the determination of the masses of the $Z$ boson
from EWSB as well as in the determination of the top quark mass. For
experimental requirements on the spectrum, one can demand for instance that it
does not lead to large radiative corrections to the precisely measured
electroweak parameters or too large values for the anomalous magnetic moment of
the muon and the branching ratio of the radiative decay of the $b$--quark. 
{\tt SuSpect} provides such tests in the form of warning or error flags in the
output file.

\subsubsection*{3.4.1 CCB and UFB}

As explained previously, in {\tt Suspect}, the EWSB conditions are
consistently  implemented by iteration on the parameters $\mu$ and $B$ and the
occurrence of a local minimum is checked numerically. In the same time one
needs to check for the non existence of deep CCB minima or UFB  directions.
Avoiding such cases may put strong constraints on the model and we mentioned in
section 3.2.3 that we have already implemented two simple CCB and UFB
conditions \cite{CCBold} and the program gives a  
warning flag when they are not
satisfied. \s

In a next version of the code, to be released in a near future, we will address 
the question of CCB  minima and UFB
directions in the most complete possible way, given the present state of the
art. Three complementary features should be considered in relation to the CCB
minima: $(i)$ the directions in the space of scalar fields along which such
minima can develop $(ii)$ whether they are lower than the EWSB minimum $(iii)$
whether  the EWSB (then false)  vacuum can still be sufficiently stable. In
Ref.~\cite{CCB}  a systematic study of point $(i)$ has been carried out
considering subspaces involving the fields $H_u, \tilde{Q}_u, \tilde{u}_R$
($H_d$ and possibly $\tilde{L}$). However, the identified D--flat directions
contain the true minima only in the case of universal scalar soft masses at the
low energy relevant scales, otherwise they constitute only {\sl sufficient}
conditions for the occurrence of CCB minima. While such directions provide very
good approximations for the first two generations, special attention should be
paid to the third generation sector as was stressed in  Ref.~\cite{lemouel}. 
This is relevant in particular to codes like \sus\ where various SUSY model
assumptions can be considered, including non--universality. Furthermore, the
check of point $(ii)$ as done in Ref.~\cite{CCB} involves a numerical scan over
field values. Actually there are cases where field--independent conditions can 
be obtained even in the case of 5--field directions $H_u, \tilde{Q}, \tilde{u}_
R, H_d, \tilde{L}$, leading to faster algorithms; see Ref.~\cite{lemouel}
and unpublished study. We will thus optimize in \sus\ the various available
complementary approaches. Point $(iii)$ has also its importance as it can
increase the phenomenologically allowed regions of the MSSM parameter space.
Some simple criteria will be encoded, following for instance 
Ref.~\cite{Kusenko}.  Finally, the UFB directions as identified in 
Ref.~\cite{CCB}, in particular UFB--3, lead to very strong constraints. 
Nonetheless, there is 
still room for some improvements by optimizing the criterion of  ``deepest
direction", leading in some cases to even stronger constraints which will be
also implemented in {\tt SuSpect}.

\subsubsection*{3.4.2 Fine--tuning}

One of the main motivations for low energy SUSY is that it solves technically
the hierarchy and naturalness problems. However, since the  $Z$ boson mass is
determined by the soft SUSY--breaking masses $M_{H_u}^2,  M_{H_u}^2$ and the
parameter $\mu^2$ as can be seen from eq.~(11), naturalness  requires that
there are no large cancellations when these parameters are  expressed in terms
of the fundamental parameters of the model [for instance  $m_0, m_{1/2}, \mu,
B$ in mSUGRA], otherwise fine--tuning is re--introduced 
\cite{RGE2,fine-tuneold,fine-tune}.  A similar problem
occurs in the case of  the top quark mass, since it is related to the top
Yukawa coupling and  $\tb$.  Various criteria for quantifying the degree of
fine--tuning in the determination of $M_Z$ and $m_t$ have been proposed and some
subjectivity is involved in the statement of how much fine tuning can be
allowed. Therefore, in our case,  we simply evaluate the sensitivity
coefficients for $M_Z^2$ and $m_t$ with respect to a given parameter $a$
\cite{fine-tune}
\beq
\frac{\delta M_Z^2}{M_Z^2} &=& C(M_Z^2, a) \, \frac{\delta a}{a} \non  \\ 
 \frac{\delta m_t}{m_t} &=& C(m_t, a) \,  \frac{\delta a}{a} 
\eeq
and leave to the user the decision of whether the amount of fine--tuning [large 
values of the $C$ coefficients] is bearable or not. We evaluate only the 
fine--tuning with respect to variations of the parameters $\mu^2$ and $B\mu$, 
for which the coefficients take the simple form: 
\beq
{\rm FT1MZ} &:&  C(M_Z^2, \mu^2) = \frac{2\mu^2}{M_Z^2} \left[ 1+ t_\beta
\frac{4 \tan^2\beta (\bar{m}_1^2-
\bar{m}_2^2) }{ (\bar{m}_1^2-\bar{m}_2^2)t_\beta-M_Z^2} \right] \non \\
{\rm FT1MZ} &:& C(M_Z^2, B\mu) = 4 t_\beta \, \tan^2\beta  \, \frac{
\bar{m}_1^2-\bar{m}_2^2 }{ M_Z^2 (\tan^2\beta -1)^2} \non \\
{\rm FT1MT} &:&  C(m_t,\mu^2) = \frac{1}{2} C(M_Z^2, \mu^2)+ \frac{2\mu^2}
{\bar{m}_1^2+\bar{m}_2^2} \frac{1}{ \tan^2\beta-1} \non \\
{\rm FT1MT} &:&  C(m_t,B\mu) = \frac{1}{2} C(M_Z^2, B\mu)+ 
\frac{1}{ 1-\tan^2\beta} 
\eeq
with $t_\beta=(\tan^2\beta+1)/(\tan^2\beta-1)$.
Further fine--tuning tests can be made, in particular with respect to the $t,b$
Yukawa couplings and are planned to be included in future versions.

\subsubsection*{3.4.3 Electroweak precision measurements} 

Loops involving Higgs and SUSY particles can contribute to electroweak
observables which have been precisely measured at LEP, SLC and the Tevatron. In
particular, the radiative corrections to the self--energies of the $W$ and $Z$
bosons, $\Pi_{WW}$ and $\Pi_{ZZ}$, might be sizable if there is a large mass
splitting between some particles belonging to the same SU(2) doublet; this can
generate a contribution which grows as the mass squared of the heaviest
particle. The dominant contributions to the electroweak observables, in
particular the $W$ boson mass and the effective mixing angle $s_W^2$, enter via
a deviation from unity of the $\rho$ parameter \cite{drho0}, which measures the
relative strength of the neutral to charged current processes at zero momentum
transfer, i.e. the breaking of the global custodial SU(2) symmetry:
\beq    
\rho = (1-\Delta \rho)^{-1} \ ; \ \Delta \rho = \Pi_{ZZ}(0)/M_Z^2 - 
\Pi_{WW}(0)/M_W^2
\eeq
Most of the MSSM contributions to the $\rho$ parameter are small, $\Delta \rho
\lsim 10^{-4}$ \cite{drhoS}. In the case of the Higgs bosons, the contributions
are logarithmic, $\sim \alpha {\rm log} (M_h/M_Z)$, and are similar to those of
the SM Higgs boson [and identical in the decoupling limit]. The chargino and
neutralino contributions are small because the only terms in the mass matrices
which could break the custodial SU(2) symmetry are proportional to $M_W$. Since
in general, first/second generation sfermions are almost degenerate in mass,
they also give very small contributions to $\Delta \rho$. Therefore, only the
third generation sfermion sector can generate sizable corrections to the $\rho$
parameter, because of left--right mixing and [in case of the stop] the SUSY 
contribution $\propto m_f^2$ leads to a potentially large splitting between the
sfermion masses. \s

We have thus calculated $\Delta \rho$ in the MSSM, taking into account only 
the contributions of  the third generation sfermions. We include full mixing 
and in the case of the stop/sbottom doublet, also the two--loop QCD corrections
due  to gluon exchange and the correction due to gluino exchange in the heavy
gluino limit,  which can increase the contribution by 30\% or so \cite{sloop}.
One may then  require the SUSY contribution not to exceed e.g. two standard 
deviations from the SM expectation~\cite{LEPrho}: $\Delta \rho ({\rm SUSY}) 
\lsim 2 \cdot 10^{-3}$.   
The precise value of  $\Delta \rho $ as calculated by \sus\ for any
choice of parameters is written in the output files.

\subsubsection*{3.4.4 The muon (g-2)} 

\nn The muon $(g-2)$ anomalous magnetic moment has been very precisely
measured and thus generally can set strong constraints on the  additional
contribution from SUSY particles. The latest experimental value is \cite{BNL}: 
\begin{eqnarray} 
(g_\mu-2) \equiv a_\mu^{\rm exp} = (11\, 659\, 208\, \pm 6) \, 10^{-10}, \ 
\end{eqnarray} 
The theoretical value predicted in the SM, including the QED, QCD and 
electroweak corrections, has some uncertainties from the determination
of the hadronic vacuum polarization contributions. 
In particular, at the time of this writing, 
the SM theoretical prediction depends quite crucially 
on whether one takes into account the 
combined measurement from $e^+e^-$ and $\tau$ decays data, or from
$e^+e^-$ data only, for the calculation of the hadronic vacuum polarization 
via a dispersion relation \cite{SMgm2}.\s

The contribution of SUSY particles to $(g_\mu-2)$ \cite{g-2old,g-2} is mainly 
due to neutralino--smuon and chargino--sneutrino loops [if no flavor violation 
is present as is the case here]. In typical models (such as mSUGRA), the
contribution of chargino--sneutrino loops usually dominates. This is also
true for configurations where all superpartners are almost degenerate in mass.
 In the latter case the contribution of $\chi_i^\pm$--$\tilde{\nu}$
loops can be approximated by [$\tilde{m}$ is the mass of the heaviest particle
per GeV]:  $\Delta a_{\mu}^{\tilde\chi^\pm \tilde{\nu}} \sim 10^{-5}
\times (\tb / \tilde{m}^2)$, to be compared with the contribution of
$\chi_i^0$--$\tilde{\mu}$ loops, $\Delta a_{\mu}^{\tilde\chi^0 \tilde{\mu}}
\sim - 10^{-6} \times (\tb / \tilde{m}^2)$.  These contributions are large for 
large values of $\tb$ and small values of the scalar and gaugino masses.
Moreover, in some other parts of the parameter space and for relatively heavy 
SUSY particles, the dominant SUSY effect follows the sign of the $\mu$ 
parameter.\s

We have included a routine which calculates the full one--loop contributions of
chargino--sneutrino and neutralino--smuon loops in the MSSM, using the
analytical expressions given in Ref.~\cite{g-2}. In this case, the full mixing
in the smuon sector is of course included [this is the only place where the
$A_\mu$ parameter plays a role in the code]. We also took into account
in this evaluation the leading two-loop QED correction\cite{gm2QED}, which
essentially reduce by about $\sim 7\%$ the one--loop contribution.\s

In view of the above mentioned present uncertainties of the SM predictions,
we refrain from providing a particular interpretation of
the comparison between the experimental above value and one of the
SM theoretical predictions, in terms of e.g. standard deviations from SM. 
Rather, the supersymmetric contributions
are simply written in  the \sus\ output files for any
choice of input parameters and models, and can be directly
compared with the (eventually updated) preferred range of the 
quantity $a^{\rm exp}_\mu -a^{\rm th,SM}_\mu$. 

\subsubsection*{3.4.5 The radiative decay $b \to s\gamma$} 

Another observable where SUSY particle contributions might be large is the
radiative flavor changing decay $b\to s\gamma$ \cite{bsg0,bsg1}. In the SM, 
this decay is mediated by loops containing charge 2/3 quarks and $W$--bosons 
but in SUSY theories, additional contributions come from loops involving 
charginos and stops or top quarks and charged Higgs bosons [contributions from
loops involving gluinos or neutralinos are very small \cite{bsg0} in the models
considered here].  Since SM and SUSY contributions appear at the same order of
perturbation theory, the measurement of the inclusive 
$B \ra X_s \gamma$ decay
branching ratio \cite{PDG}
is a very powerful tool for
constraining  the SUSY parameter space. \s

In Refs.~\cite{bsg1,paolo},  the next--to--leading order QCD corrections to 
the decay rate in the MSSM have been calculated and a {\sc Fortran} code 
has been provided. It gives the most up--to--date determination of BR($b \to 
s\gamma)$ where all known perturbative and non--perturbative effects are 
implemented, including all the possibly large contributions which can occur at 
NLO, such as terms $\propto \tan \beta$ and/or terms containing
logarithms of $M_{\rm EWSB}/M_W$. We have interfaced this routine with our
code\footnote{We thank Paolo Gambino for providing us with his code and for his
help in interfacing it with ours.}.  Besides the fermion and gauge boson masses
and the gauge couplings that we have as inputs, we will use the values of the
other SM input parameters required for the calculation of the rate given in
Ref.~\cite{bsg3}, except for the cut--off on the photon energy, $E_{\gamma} >
(1-\delta)m_b/2$ in the bremsstrahlung process $b \to s\gamma g$, which we fix
to $\delta=0.9$ as in Ref.~\cite{paolo}.  Then, again, the theoretical value of 
${\rm BR}(b \ra s \gamma) $ as calculated by the above mentioned routine is
written in the output files, and can be compared
by the user with the latest available data \cite{PDG}.

\section*{4. Running {\tt SuSpect}}

\subsection*{4.1 Basic facts about {\tt SuSpect}}
The program {\tt Suspect} is composed of several files and routines: \s

$i)$ \underline{The input files {\tt suspect2.in}} (standard format) or
alternatively \underline{\tt suspect2$\_$lha.in} (the SLHA format): here one
can select the model to be investigated, the accuracy of the algorithm, the
input data (SM fermion masses and gauge couplings). Some reasonable default
values are set in the example which is provided.  One would then simply select
the SUSY model (pMSSM, mSUGRA, GMSB and AMSB), choose the corresponding input
parameters and possibly make a few choices concerning the physical calculation
(such as enforcing or not unification of the gauge couplings, changing the
scale at which EWSB occurs, including or not RC to the masses,
and choosing the routine calculating the Higgs masses). 
The list of possible choices is given
in the next subsection. \s 

$ii)$ \underline{The program {\tt suspect2\_call.f}}: this is an example of a
routine which calls the main subroutine {\tt suspect2.f}. This program is
necessary to run \sus\ since it defines the primary algorithm control input
parameters needed by the latter. In particular, there is an important parameter
({\tt INPUT}) which controls the form of input/output: depending on its value
it allows to select the input and output either in the standard  ({\tt
suspect2.in}) or in the SLHA  ({\tt suspect2$\_$lha.in}) format, or
alternatively to bypass the reading of any input files, in which case all the
parameters and choices are to be defined by the user in this calling routine.
This last choice is particularly useful for interfacing  \sus\ with other
routines and/or for scans of the parameter space. 

$iii)$ \underline{The main routine {\tt suspect2.f}}: here all the calculation
of the spectrum is performed, once the input is supplied by {\tt suspect2.in},
{\tt suspect2$\_$lha.in} or {\tt suspect2$\_$call.f}. This routine is
self--contained, except for the determination of the Higgs boson masses where
it needs to call one additional routine for the two--loop contributions that we
also supply: {\tt twoloophiggs.f} (originally provided by Pietro Slavich). 
Note that we provide also as a separate file, needed in the compilation: {\tt
bsg.f}, performing the calculation of the $b \to s \gamma$ branching ratio
(originally provided by Paolo Gambino).

$iv)$ \underline{output files {\tt suspect2.out} and {\tt
suspect2$\_$lha.out}}: these files are generated by default [they can be
switched off by an appropriate value of the control parameter {\tt INPUT} in
{\tt suspect2\_call.f}] at each run of the program and gives the results for
the output soft SUSY--breaking parameters [when they are calculated] and the
masses and mixing angles of the Higgs and SUSY particles.  Some warnings 
and comments are also given when  the obtained spectrum is problematic as will 
be discussed in the next subsection.  \s 

The routine {\tt suspect2.f} consists of about 10.000 lines of code and takes 
about 350 Ko of memory, while the input and the calling routines have only a 
few hundred lines (most of them being comments). The accompanying 
separate routine for the calculation of the two-loop radiative correction
to the Higgs masses has about 7000 lines of code.  
The complete executable code takes about 0.9 Mo space. \s

The {\sc Fortran} files have to be compiled altogether and, running for instance 
on a PC using {\sc gnu Fortran}, the (minimal options) 
compilation and link commands are:

g77 -c suspect2\_call.f suspect2.f twoloophiggs.f bsg.f

g77 -o suspect suspect2\_call.o suspect2.o twoloophiggs.o bsg.o

suspect\s

\nn  no other compilation option is in principle needed, though some users
might find it useful to try other standard compilation options.  The running 
time for a typical model point, for instance an mSUGRA point, 
is about 1 second on a PC with a 1 GHz processor.

\subsection*{4.2 The main routine and the control parameters}

In this subsection, while we will refrain from exhibiting the input and output 
files (some examples can be found on the web page of the program), we mention
a few important features about the main routine and the control parameters
that it is uses. 
\begin{verbatim}
         SUBROUTINE SUSPECT2(iknowl,input,ichoice,errmess)
\end{verbatim}
is the main routine of the program, to be used as it is or to be
called by any other routine (such as {\tt suspect2\_call.f}, as will be
discussed below). It has the following four basic input control 
parameters: \s

{\tt IKNOWL}: which sets the degree of control on the various parts of the 
algorithm. It has two possible values: 

\begin{itemize}
\vspace*{-3mm}
\item[--] {\tt IKNOWL=0}: blind use of the program, i.e. no control on
any  ``algorithmic" parameter. Reasonable
default values are set for the control parameters and the program gives just
the results from the physical input.  
\vspace*{-3mm}

\item[--]  {\tt IKNOWL=1}: in addition 
 warning/error messages are collected in the {\tt suspect2.out} 
file (this is the recommended choice in general).  
\vspace*{-3mm}

\end{itemize}

{\tt INPUT}: is for the physical input setting and can take the
following values:

\begin{itemize}
\vspace*{-3mm}
\item[--] {\tt INPUT=0}: the model and option parameters {\tt ichoice(1)-(10)} 
as well as the values of the input parameters are read off from the
file  {\tt suspect2.in} in the original SuSpect format.  The output is written
 in both {\tt suspect2.out} and SLHA format {\tt suspect2$\_$lha.out} files.

\vspace*{-3mm}

\item[--] {\tt  INPUT=1}: the user defines all the relevant
input choices and parameters within his calling program, i.e. 
any reading of input file(s) is bypassed. The required list of
parameters to be defined (with consistent names etc), can be found in the
commons given below. In addition, examples of input
parameter setting for this option, with all the necessary
input parameters (appearing in common blocks), are given for various
models in the calling program {\tt suspect2$\_$call.f}.
The output is written
 in both {\tt suspect2.out} and SLHA format {\tt suspect2$\_$lha.out} files.
\vspace*{-3mm}

\item[--] {\tt INPUT=2}: the model and option parameters {\tt ichoice(1)-(10)} 
as well as the values of the physical input parameters are read off from the
file  {\tt suspect2$\_$lha.in} in the SLHA format. 
The output is written
 in both {\tt suspect2.out} and SLHA format {\tt suspect2$\_$lha.out} files.
\vspace*{-3mm}

\item[--] {\tt INPUT=11}: same as {\tt  INPUT=1}, but with no output file(s) 
written (note this option may be more  convenient e.g. for scans of the 
MSSM parameter space).
\vspace*{-3mm}
\end{itemize}   

{\tt ICHOICE}: initializes the various options for 
the models to be considered,
the degree of accuracy to be required, the features to be included, etc. There
are 11 possible choices at present and the options are described in detail in
the input files  {\tt suspect2.in} or {\tt suspect2$\_$lha.in}: 
\begin{itemize}
\vspace*{-3mm}
\item[--] {\tt ICHOICE(1)}: Choice of the model to be considered.
\vspace*{-3mm}

\item[--] {\tt ICHOICE(2)}: For the perturbative order (1 or 2 loop) of the 
RGEs. 
\vspace*{-3mm}

\item[--] {\tt ICHOICE(3)}: To impose or not the GUT scale. 
\vspace*{-3mm}

\item[--] {\tt ICHOICE(4)}: For the accuracy of the RGEs.
\vspace*{-3mm}

\item[--] {\tt ICHOICE(5)}: To impose or not the radiative EWSB. 
\vspace*{-3mm}

\item[--] {\tt ICHOICE(6)}: To chose different input in general MSSM.
\vspace*{-3mm}

\item[--] {\tt ICHOICE(7)}: For the radiative corrections to the (s)particles 
masses. 
\vspace*{-3mm}

\item[--] {\tt ICHOICE(8)}: To set the value of the EWSB scale.
\vspace*{-3mm}

\item[--] {\tt ICHOICE(9)}: For the final accuracy in the spectrum 
calculation (two possibilities).
\vspace*{-3mm}

\item[--] {\tt ICHOICE(10)}: For the approximation in
 calculating the Higgs boson masses (approximate, one-loop, or two-loop).
\vspace*{-2mm}
\end{itemize}

{\tt  ERRMESS}: which provides a useful set of warning/error message flags,
           that are automatically written in the output files
 {\tt suspect2.out} or  {\tt suspect2$\_$lha.out}:
\begin{itemize}
\vspace*{-3mm}
\item[--] {\tt ERRMESS(i)= 0}: Everything is fine,
\vspace*{-3mm}

\item[--] {\tt ERRMESS(1)=-1}: tachyonic 3rd generation sfermion from RGE,
\vspace*{-3mm}

\item[--] {\tt ERRMESS(2)=-1}: tachyonic 1st/2d generation sfermion from RGE,
\vspace*{-3mm}

\item[--] {\tt ERRMESS(3)=-1}: tachyonic $A$ boson (maybe temporary: see final 
mass), 
\vspace*{-3mm}

\item[--] {\tt ERRMESS(4)=-1}: tachyonic 3rd generation sfermion from mixing,
\vspace*{-3mm}

\item[--] {\tt ERRMESS(5)=-1}: $\mu(M_{\rm GUT})$ guess inconsistent, 
\vspace*{-3mm}

\item[--] {\tt ERRMESS(6)=-1}: non--convergent $\mu$ from EWSB, 
\vspace*{-3mm}

\item[--] {\tt ERRMESS(7)=-1}: EWSB maybe inconsistent  (but RG--improved only
check),
\vspace*{-3mm}

\item[--] {\tt ERRMESS(8)=-1}: $V_{\rm Higgs}$ maybe UFB or CCB (but 
RG--improved only check),
\vspace*{-3mm}

\item[--] {\tt ERRMESS(9)=-1}: Higgs boson masses are NaN, 
\vspace*{-3mm}

\item[--] {\tt ERRMESS(10)=-1}: RGE problems (non--perturbative and/or Landau 
poles).
\end{itemize}

\subsection*{4.3. Calculations with {\tt SuSpect}}

\subsubsection*{4.3.1 Comparison with other codes} 

Our results for some representative points of the MSSM parameter space have
been carefully cross--checked against other existing codes. Most of the earlier
comparisons \cite{comp} have been performed in the context of mSUGRA type
models but, more recently,  comparisons in other models have been performed.  
We obtain, in general, a very good agreement, at the percent level and often 
better \cite{compsabi}, with the codes \softsusy\ and \spheno\footnote{We thank
Ben Allanach, Sabine Kraml, Werner Porod and Pietro Slavich for their gracious 
help in performing these detailed comparisons of the programs.}.  The most
sophisticated parameter to obtain in this context is the lightest Higgs boson
mass, since it incorporates all possible ingredients: the RGE's for the
evaluation of $M_{H_u}$ and $M_{H_d}$, the effective potential and the EWSB for
the determination of $M_A$ and the tadpoles, the radiative corrections to the
Higgs sector which involve also the two--loop corrections, etc.  Recently, a
detailed comparison of \sus\ with   \softsusy\ and \spheno\ (and also FeynHiggs)
for the Higgs boson masses in many scenarios (mSUGRA, GMSB, AMSB, general MSSM)
was performed in Ref. \cite{adkps}. The value that we obtain for $M_h$, for
instance, is in a very good agreement, the difference being less than half a
GeV, while the agreement for $M_A$ and $\mu$ is at the level of 1\%; see
Ref.~\cite{adkps} for details. The agreement with the program ${\tt ISASUGRA}$
is not as good, in particular for what concerns the Higgs sector, since
different approximations have been used in the two cases.  

\subsubsection*{4.3.2 Interface with other programs}

In the way it is written, \sus\ can be easily interfaced with other programs or
Monte--Carlo event generators\footnote{To make this interfacing easier, in
addition to including the input and output files in the now standard SLHA
format, we also have provided a set of obvious commons for the input and output
parameters needed or calculated by \sus\ and named all commons, subroutines and
functions used with a prefix {\tt SU\_}, in order not to be in conflict with
those used by other programs.}.  Private or official versions of some programs 
exist in which an interface with \sus\ has been made; we give a short list of 
them\footnote{Note that
we have also interfaced the program with a private code written by Manuel Drees
calculating the cosmological relic density of the lightest neutralinos for the
complete analysis of the mSUGRA parameter space performed in Ref.~\cite{DDK}.}:

\begin{itemize}
\vspace*{-2mm}
\item {\tt micrOMEGAs} \cite{micromegas}: for the automatic (analytical and then
numerical)  calculation of the cosmological relic density of the lightest 
neutralinos, including all possible channels.
\vspace*{-2mm}

\item {\tt DARKSUSY} \cite{darksusy}: also for the calculation of the relic 
density of the lightest neutralinos [including co--annihilation effects] and 
their direct and indirect detection  rates as well as other astrophysical
features. [The program has its own calculation of the SUSY spectrum, but it is 
rather approximate]\footnote{Note that a private {\tt Suspect/DARKSUSY} 
interface has been already used for prediction studies of indirect
LSP detection \cite{nezri,falvard}.} . 
\vspace*{-2mm} 

\item {\tt HDECAY} \cite{HDECAY}: for the calculation of the decay branching 
ratios and total decay widths of the SM and MSSM Higgs bosons [in fact some 
routines, in particular those for the QCD running and for the interface with 
the  routines calculating the Higgs boson masses, are borrowed from there].
\vspace*{-2mm}

\item  {\tt SDECAY} \cite{SDECAY}: for the calculation of the decay widths and 
branching ratios of  SUSY particles including higher order [three--body decays 
for gauginos and stops, four--body decays for the lightest stop and QCD 
corrections to the two--body decays of squarks and gluinos], which will 
appear soon.  
\vspace*{-2mm}

\item {\tt SUSYGEN} \cite{SUSYGEN}: a Monte--Carlo event generator for Higgs 
and SUSY particle production in the MSSM [mainly in $e^+ e^-$ collisions
but some processes in $ep$ and $pp$ collisions are implemented]. The program
is also interfaced with {\tt HDECAY}. 
\vspace*{-2mm}

\item {\tt SFITTER} \cite{SFITTER} which determines the weak scale MSSM 
parameters from measurements performed at hadron and $e^+e^-$ colliders from
a combined fit/grid taking into account the correlations.  
\vspace*{-2mm}
\end{itemize}

As discussed already, we also interfaced \sus\  with  the code {\tt bsg.f}
which calculate the rate of the $b \to s\gamma$ decay at next--to--leading 
order \cite{paolo}.  An interface with the Monte--Carlo event generators 
such as {\tt PYTHIA} \cite{pythia} and {\tt HERWIG} \cite{herwig} can be
made throgh the SLHA format. 

\subsubsection*{4.3.3 Web information and maintenance}

A web page devoted to the \sus\ program can be found at the http address: \s

\centerline{\tt http://www.lpta.univ-montp2.fr/\~\,kneur/Suspect}\s 

\nn It contains all the information that one needs on the program: 

-- Short explanations of the code and how to run it. 

-- The complete ``users manual" can be obtained in post-script or PDF form.

-- A list of important changes/corrected bugs in the
code. 

-- A mailing list to which one can subscribe to be automatically advised about 
future\\ \hspace*{.8cm} \sus\  updates  or eventual corrections. \s

\nn One can also download directly the various files of the program: 

-- {\tt suspect2.in},  {\tt suspect2$\_$lha.in}: the input files in standard
or SLHA format.

-- {\tt suspect2$\_$call.f}: the calling program sample.

-- {\tt suspect2.f}: the main routine of the program.

-- {\tt twoloophiggs.f}: the additional Higgs routine needed.

-- {\tt SuSpect2.tar.gz}: all needed routines in standard compressed format 
for the latest version.

-- {\tt SuSpect2$\_$Old.uu}: the routines for the previous versions of the 
program. \s

\nn Note that the possibility to use directly the program [as well as the 
other RGE codes] interactively on the web has been set by Sabine Kraml 
\cite{compsabi}. 

\section*{5. Conclusion} 

We have presented the version 2.3 of the {\sc Fortran} code {\tt SuSpect}  which
calculates the Supersymmetric and Higgs particle spectrum in the  MSSM. The
calculation can be performed in constrained models  with universal boundary
conditions at high scales such as the gravity (mSUGRA), anomaly (AMSB) or gauge
(GMSB) mediated breaking models, but also in the  non--universal or
unconstrained MSSM case, with up to 22 free input parameters which can be set
either at the electroweak symmetry breaking scale or obtained from boundary
conditions on some common parameters at a high--energy scale. \s

A particular care has been taken to treat all the mandatory features which
are needed to describe accurately these various scenarii: the renormalization
group evolution of parameters between low and  high energy scales, the
consistent implementation of radiative electroweak  symmetry breaking and the
calculation of the physical masses of the Higgs  bosons and supersymmetric
particles taking into account all dominant radiative corrections. The program
provides several  options [for accuracy, scale choice, etc...] to deal with
these aspects. \s  

The program can check the fulfillment of theoretical
constraints, such as the absence of tachyonic particles and improper lightest
SUSY particle, the absence of non desired charge and color breaking as well as
unbounded from below minima and a large fine--tuning in the electroweak
symmetry breaking conditions. A verification that the obtained spectrum is in
agreement with high precision measurements such as the $\rho$ parameter, the 
muon $g-2$ and the radiative $b \to s\gamma$ decay, can also be performed.\s

The  program has a high degree of flexibility in the choice of the model and/or
the input parameters and an adequate level of approximation at different
stages. It is rather precise and quite reliable [since it has been compared
with several other similar existing codes], relatively fast to allow for rapid
comprehensive scans of the parameter space and simple enough to be linked with
 other programs dealing with MSSM particle properties or with Monte--Carlo
event generators. We have also
provided a simpler way to run the code interactively on the web in the
constrained models. \s

Several upgrades, which include the
possibility to analyze additional theoretical models and to 
further extend the already existing interfaces
with  programs for (s)particle decay branching ratios and Dark Matter
calculations and with some Monte--Carlo event generators to simulate the
production properties, are planned.  

\subsection*{Acknowledgments:}
In its latest present version, the code \sus\ benefited largely
from the invaluable help and many cross--checks of Pietro Slavich, 
in particular, in implementing a consistent interface with his
routine for the $\drbar$ calculation
of the Higgs masses including the leading two--loop contributions.   
In former stages the code has been developed in 
the framework of the French 
working group (GDR--SUSY), organized by the {\it 
Centre National de la Recherche Scientifique} (CNRS), and has
 been checked and
``debugged" during the last few years with the help of several members of the
``MSSM" and ``Tools" working groups of the GDR to whom we are indebted. We
thank in particular, Genevieve B\'elanger, 
Fawzi  Boudjema,
Marie-Bernadette Causse, Jean-Baptiste de Vivie, Laurent Duflot, Nabil
Ghodbane, Jean-Francois Grivaz, Cyril Hugonie, Stavros Katsanevas, Vincent
Lafage, R\'emi Lafaye, 
Imad Laktineh, Christophe Le Mou\"el,  Yann Mambrini, Steve Muanza,
Margarete M\"uhlleitner,  Emmanuel Nezri, Jean Orloff, Emmanuelle Perez, Sylvie
Rosier--Lees, Roberto Ruiz de Austri, Dirk Zerwas. 
We have as well similarly benefited from the working group ``SUSY
Dark Matter" of the GDR {\sl Ph\'enom\`enes Cosmiques de Hautes Energies}. We
thank all members of this group, in particular Julien Guy, Agnieszka 
Jacholkowska, Julien Lavalle, Eric Nuss and Mariusz Sapinski for various 
cross-checks.  
We have also benefited from several discussions, comments and help from: Ben
Allanach, Pierre Bin\'etruy, Andreas Birkedal, Francesca Borzumati, Aseshkrishna
Datta, Manuel
Drees, Paolo Gambino, Naveen Gaur, Sven Heinemeyer, Jan Kalinowski, Yeong Gyun
Kim, Sabine Kraml, Filip Moortgat, Stefano Moretti, Takeshi Nihei, Werner
Porod, Fran\c cois Richard, Peter Richardson, Leszek Roszkowski, Aurore
Savoy--Navarro, 
Peter Skands, Michael Spira and Charling Tao. 
We thank  them all. Finally, thanks
to our system--manager in Montpellier, Dominique
Caron, for technical help.



\begin{thebibliography}{99} 
%
\bibitem{SUSY} For an introduction, see: ``Supersymmetry and
Supergravity" by J. Wess and J. Bagger, Princeton Series in Physics.
%
\bibitem{MSSM} For reviews on the MSSM, see: 
H.P. Nilles, Phys. Rep. 110 (1984) 1;
R. Barbieri, Riv. Nuov. Cim. 11 (1988) 1; 
R. Arnowitt and Pran Nath, Report CTP-TAMU-52-93; 
M. Drees and S.P. Martin, hep-ph/9504324; 
J. Bagger, Lectures at TASI-95, hep-ph/9604232; 
S.P. Martin, hep-ph/9709356; 
S. Dawson, hep-ph/9712464. 
%
\bibitem{Haber-Kane} H. E. Haber and G. Kane, Phys. Rep. 117 (1985) 75;
J.F. Gunion and H.E. Haber, Nucl. Phys. B272 (1986) 1 and Nucl. Phys. B278
(1986) 449 ; (E) hep-ph/9301205. 
%
\bibitem{GDR} A. Djouadi and S. Rosiers--Lees {\sl et al}, Summary Report of the 
MSSM Working Group for the ``GDR--Supersym\'etrie", hep-ph/9901246.
%
\bibitem{HHG} For a review on the Higgs sector of the MSSM, see J.F. Gunion, 
H.E. Haber, G.L. Kane and S. Dawson, ``The Higgs Hunter's Guide", 
Addison--Wesley, Reading 1990.
%
\bibitem{searches} M. Carena et al., Report of the Higgs WG hep-ph/0010338 and
S. Abel et al, Report of the MSSM WG  hep-ph/0003154, for ``RUN II at the 
Tevatron";  CMS Coll., Technical Proposal, report CERN/LHCC/94-38 (1994);
ATLAS Coll., Technical Design Report, CERN/LHCC/99-15 (1999); Proceedings 
of the Les Houches Workshops: A. Djouadi et al., hep-ph/0002258 (1999) and 
D. Cavalli et al., hep-ph/0203056 (2001); 
E. Accomando, Phys. Rept. 299 (1998) 1;  
American Linear Collider Working Group,  hep-ex/0106057; 
ACFA Linear Collider Working group hep-ph/0109166; 
TESLA Technical Design Report, Part III,  hep-ph/0106315.  
%
\bibitem{parameters} S. Dimopoulos and D. Sutter, Nucl. Phys. B452 (1995) 496; 
see also the discussions given by H.E. Haber, hep-ph/9709450 and G.L. Kane, 
hep-ph/0008190. 
%
\bibitem{mSUGRA} 
A.H. Chamseddine, R. Arnowitt and P. Nath, Phys. Rev. Lett. 49 (1982) 970;
R. Barbieri, S. Ferrara and C.A Savoy, Phys. Lett. B119 (1982) 343;
L. Hall, J. Lykken and S. Weinberg, Phys. Rev. D27 (1983) 2359.
%
\bibitem{AMSB} L. Randall and R. Sundrum, Nucl. Phys. B557 (1999) 79;
G. Giudice, M. Luty, H. Murayama and R. Rattazzi, JHEP 9812 (1998) 027.
%
\bibitem{AMSB1} J.A. Bagger, T. Moroi and E. Poppitz, JHEP 0004 (2000) 009.

\bibitem{AMSBp} For a recent phenomenological review of AMSB models, see for
instance: K. Huitu, J. Laamanen and P. N. Pandita, Phys.~Rev. D65 (2002) 115003.
%
\bibitem{GMSB} For a general review: G.F. Giudice and R. Rattazzi, 
Phys.~Rept.~322 (1999) 419.
%
\bibitem{GMSBp} For phenomenological reviews of GMSB models, see for instance: 
S. Ambrosanio, G.D. Kribs and S.P. Martin, Phys. Rev. D56 (1997) 1761; 
C.H. Chen and J.F. Gunion, Phys. Rev. D58 (1998) 075005.
%
\bibitem{ISASUGRA} H. Baer, F.E. Paige S.D. Protopopescu and X. Tata, ``ISAJET",
hep-ph/0001086. 
%
\bibitem{SOFTSUSY}  B.C. Allanach, ``SOFTSUSY", Comput. Phys. Commun. (2002) 
143. 
%
\bibitem{SPHENO} W. Porod, ``SPHENO", Comput.\ Phys.\ Commun. 153 (2003) 275.
%
\bibitem{SUSPECT} A preliminary version of the program exists since 1998
and was described in A. Djouadi, J.L. Kneur and G. Moultaka in Ref.~\cite{GDR}.
\bibitem{RGE} K. Inoue, A. Kakuto, H. Komatsu and S. Takeshita,
Prog. Theor. Phys. 68, 927 (1982); Erratum: ibid. 70, 330 (1983);
ibid. 71, 413 (1984); M. Machacek and M.T. Vaughn, Nucl. Phys. B222 (1983) 
83; ibid. B236 (1984) 221; ibid B249 (1985) 70; I. Jack, Phys. Lett. B147 
(1984) 405; W. de Boer, R. Ehret and D.I. Kazakov, Z. Phys. C67 (1994) 647;
Y. Yamada, Phys. Rev. D50 (1994) 3537; I. Jack and D.R.T. Jones, Phys. Lett.
B333 (1994) 372. 
%
\bibitem{RGE2} 
D.J. Casta\~no, E.J. Piard and P. Ramond, Phys. Rev. D49 (1994) 4882;
V. Barger, M.S. Berger and P. Ohmann, Phys. Rev. D47 (1993) 1093.
%
\bibitem{drbar} 
I. Jack, D. Jones, S. Martin, M. Vaughn and Y. Yamada, Phys. Rev. D50 (1994) 
5481; S. Martin and M. Vaughn, Phys. Lett. B318, 331 (1993) and Phys. Rev. 
D50 (1994) 47.
\bibitem{EWSB} K. Inoue et al. in Ref.~\cite{RGE}, L. Iba\~nez and G.G. Ross 
Phys. Lett. B110 (1982) 227; L. Alvarez--Gaum\'e, J. Polchinski and M.B. Wise, 
Nucl. Phys. B221 (1983) 495; J. Ellis, J. Hagelin,
D. Nanopoulos and K. Tamkavis, Phys. Lett. B125 (1983) 275, L.E. Ibanez and 
C. Lopez, Nucl. Phys. B233 (1984) 511; L.E. Ibanez,  C. Lopez and C. Munoz, 
Nucl. Phys. B250 (1985) 218. 
%
\bibitem{potential} 
R. Arnowitt and P. Nath, Phys. Rev. D46, 3981 (1992); 
V. Barger, M.S. Berger and P. Ohmann, Phys. Rev. D49, 4908 (1994); 
P.H. Chankowski, S. Pokorski and J. Rosiek, Nucl. Phys. B423 (1994) 437. 
%
\bibitem{Vscale} 
G. Gamberini, G. Ridolfi and F. Zwirner, Nucl. Phys. B331, 331 (1990);
B. de Carlos and J.A. Casas, Phys. Lett. B309, 320 (1993). 
%
\bibitem{Spanos} A.B. Lahanas and V.C. Spanos, Eur. Phys.~J. C23 (2002) 185.
\bibitem{PBMZ} 
D.M. Pierce, J.A. Bagger, K. Matchev and R.J. Zhang, Nucl. Phys. B491
(1997) 3.
%
\bibitem{RC0} For some earlier work on SUSY radiative corrections is for
instance: J.A.  Grifols and J. Sola, Phys. Lett. B137 (1984) 257 and Nucl.
Phys. B253 (1985) 47; P. Chankowski et al., Nucl. Phys. B417 (1994) 101; D.
Pierce and A.  Papadopoulos, Nucl. Phys. B430 (1994) 278 and Phys. Rev. D50
(1994) 565; A.  Donini, Nucl. Phys. B467 (1996) 3; N. Krasnikov, Phys. Lett.
B345 (1995) 25; A.B. Lahanas, K. Tamkavis and N.D. Tracas, Phys. Lett. B324
(1994) 387; M. Drees, M.M. Nojiri, D.P. Roy and Y. Yamada, Phys. Rev. D56
(1997) 276.  
%
\bibitem{RCH} For radiative corrections in the Higgs sector, see for instance: 
Y. Okada, M. Yamaguchi and T. Yanagida, Prog. Theor. Phys. 85 (1991) 1; 
H. Haber and R. Hempfling, Phys. Rev. Lett. 66 (1991) 1815; 
J. Ellis, G. Ridolfi and F. Zwirner, Phys. Lett. 257B (1991) 83;
R. Barbieri, F. Caravaglios and M. Frigeni, Phys. Lett. 258B (1991) 167;
M.~Drees and M.M.~Nojiri, Phys. Rev. D45 (1992) 2482;
R. Hempfling and A. Hoang, Phys. Lett. B331 (1994) 99;
M. Carena, J. Espinosa, M. Quiros and C. Wagner, Phys. Lett. B335 (1995) 209;
M. Carena, M. Quiros and C.E.M. Wagner, Nucl. Phys. B461 (1996) 407;
H. Haber, R. Hempfling and A. Hoang, Z. Phys. C75 (1997) 539;  
S. Heinemeyer, W. Hollik and G. Weiglein, Phys. Rev. D58 (1998) 091701;
J.R. Espinosa and R.J. Zhang, JHEP0003 (2000) 026 and Nucl. Phys. B586 (2000)3;
M. Carena et al.,  Nucl. Phys. B580 (2000) 29. 
%
\bibitem{FHF} S. Heinemeyer, W. Hollik and G. Weiglein, Comput.\,Phys.\,Commun.
124 (2000) 76. 
%
\bibitem{FeynHiggsFast} S. Heinemeyer, W. Hollik and G. Weiglein, 
hep-ph/0002213. 
%
\bibitem{dsz}
G.~Degrassi, P.~Slavich and F.~Zwirner,
Nucl.\ Phys.\ B611 (2001) 403. 
%
\bibitem{bdsz}
A.~Brignole, G.~Degrassi, P.~Slavich and F.~Zwirner,
Nucl.\ Phys.\ B631 (2002) 195.
%
\bibitem{bdsz2}
A.~Brignole, G.~Degrassi, P.~Slavich and F.~Zwirner,
Nucl.\ Phys.\ B643 (2002) 79.
%
\bibitem{dds}
A.~Dedes, G.~Degrassi and P.~Slavich, 
Nucl. \ Phys. \ B672 (2003) 144.
%
\bibitem{adkps} B.C.~Allanach, A.~Djouadi, J.-L.~Kneur, W.~Porod
and P.~Slavich, JHEP 0409 (2004) 044.
%
\bibitem{slha}
P.~Skands {\em et al.}, JHEP 0407 (2004) 036. 
%
\bibitem{MSSMdef} P. Fayet, Phys. Lett. 69B (1977) 489; G. Farrar and P. Fayet, 
Phys. Lett. 76B (1978) 575; N. Sakai, Z. Phys. C11 (1981) 153; S. Dimopoulos 
and H. Georgi, Nucl. Phys. B193 (1981) 150; K. Inoue, A. Komatsu and S. 
Takeshita, Prog. Theor. Phys 68 (1982) 927; E. Witten, Nucl. Phys. B231 (1984) 
419. 
%
\bibitem{DF} P. Fayet and  J. Iliopoulos, Phys. Lett. 51B (1974) 461;
L. O'Raifeartaigh, Nucl. Phys. B96 (1975) 331. 
%
\bibitem{soft} L. Girardello and M.T. Grisaru, Nucl. Phys. B194 (1982) 65.
%
\bibitem{flavorev} For a review, see: A. Masiero and  L. Silvestrini, 
hep-ph/9711401. 
%
\bibitem{CCBold} J.M. Frere, D.R.T. Jones and S. Raby, Nucl. Phys. B222 (1983)
11; L. Alvarez--Gaum\'e, J. Polchinski and M.B. Wise, Nucl. Phys. B221 (1983) 
495; J.P. Derendiger and C.A. Savoy, Nucl. Phys. B237 (1984) 307; C. Kounas,
A.B. Lahanas and M. Quiros, Nucl. Phys. B236 (1984) 438; M. Claudson, L.J. Hall
and I. Hinchliffe, Nucl. Phys. B236 (1983) 501.  
%
\bibitem{PDG} Particle Data Group, S. Eidelman et al, Phys. Lett. B592, 1
(2004).
%
\bibitem{LEPunif} J. Ellis , S. Kelley and D.V. Nanopoulos, Phys. Lett. 
B260 (1991) 131; U. Amaldi, W. de Boer and H. F\"urstenau, Phys. Lett. B260 
(1991) 447; P. Langacker and M. Luo, Phys. Rev.D 44 (1991) 817;
C. Giunti, C.W. Kim and U.W. Lee, Mod. Phys. Lett. A6 (1991) 1745.
%
%
%
\bibitem{AM1} T. Gherghetta, G.F. Giudice, J.D. Wells, Nucl. Phys. B559 (1999) 
27.
%
\bibitem{AM2} See e.g., D.E. Kaplan and G.D. Kribs, JHEP 09 (2000) 048.
%
\bibitem{AM3} See e.g.,  M. Carena, K. Huitu and T. Kobayashi, Nucl. Phys. 
B592 (2001) 164. 
%
\bibitem{gmsb1} M.~Dine and A.~E.~Nelson, Phys. Rev. D48 (1993) 1277;
    M.~Dine, A.~E.~Nelson and Y.~Shirman, Phys. Rev. D51 (1995) 1362;
    M.~Dine, A.~E.~Nelson, Y.~Nir and Y.~Shirman, Phys. Rev. D53 (1996) 2658.
 %
\bibitem{gmsb2} N. Arkani-Hamed, J. March-Russell and H. Murayama, Nucl. Phys. 
B509 (1998) 3; H. Murayama, Phys. Rev. Lett. 79 (1997) 18;
K.I. Izawa, Y. Nomura, K. Tobe and T. Yanagida, Phys. Rev. D56 (1997) 2886;
M.A. Luty, Phys. Lett. B414 (1997) 71. 
%
\bibitem{barbieri} R.~Barbieri, G.~Dvali and A.~Strumia, Phys. Lett. B333 
(1994) 79; E.~Witten, hep-ph/0201018. 
%
\bibitem{agashe} K.~Agashe and M.~Graesser, Nucl. Phys. B507 (1997) 34.
%
\bibitem{weldon} S.K.~Soni and H.A.~Weldon, Phys. Lett. 126B (1983) 215.
%
\bibitem{ibanez} L.E.~Ib\'a\~nez and D.~Lust, Nucl. Phys. B382 (1992) 305.
%
\bibitem{Pierre} See e.g., P. Binetruy, M.K. Gaillard and B.D. Nelson, 
Nucl.~Phys. B604 (2001) 32. 
%
\bibitem{NM1} For phenomenological discussions and references
on the original models, see for instance: G. Anderson et al., hep-ph/9609457  
and H. Baer et al., JHEP 0205 (2002) 061.
%
\bibitem{NM2} See e.g., G. Cohen, D.B. Kaplan and A.E. Nelson, Phys. Lett. B388 
(1996) 588; S. Dimopoulos and G.F. Giudice, Phys. Lett. B357 (1995) 573; 
A. Pomarol and D. Tommasini,  Nucl. Phys. B466(1996) 3; P. Binetruy and E. 
Dudas, Phys. Lett. B389 (1996) 503; B. de Carlos, J. Casas, F. Quevedo and 
E. Roulet, Phys. Lett. B318 (1993) 447; J. Bagger, J.L. Feng and N. Polonsky, 
Nucl. Phys. B563 (1999) 3; J. Bagger, J.L. Feng, N. Polonsky and R.J. Zhang, 
Phys. Lett. B473 (2000) 264; V.D. Barger, C. Kao and  R.J. Zhang, Phys. Lett. 
B483 (2000) 184.
%
\bibitem{manuel} M.~Drees, Phys. Lett. B181 (1986) 279.
%
\bibitem{KM} C.~Kolda and S.~Martin, Phys. Rev. D53 (1996) 3871.
%
\bibitem{NM3} D. Matalliotakis and H.P. Nilles, Nucl. Phys. B435 (1995) 115;
   M. Olechowski and S. Pokorski, Phys. Lett. B344 (1995) 201; 
   N. Polonsky and A. Pomarol, Phys. Rev. D51 (1995) 6352; 
   V. Berenzenski et al., Astropart. Phys. 5 (1996) 1;
   R. Arnowitt and P. Nath, Phys. Rev. D56 (1997) 2820 and 2833.
%
\bibitem{Ex-nonuniv} Example of analyses where Higgs and sfermion masses 
are disconnected are: G. Belanger et al., Nucl. 
Phys. B581 (2000) 3; A. Datta et al., Phys. Rev. D 65 (2002) 015007;
A. Djouadi et al., Phys.\ Lett.\ B376 (1996) 220 (1996)
and Eur.\ Phys.\ J.\ C1 (1998) 149. 
%
\bibitem{DM-nonuniv} V.D. Barger, F. Halzen, D.Hooper and Chung Kao, Phys.
Rev. D65 (2002) 075022; V. Bertin, E. Nezri and J. Orloff JHEP 0302 (2003) 046;
J. Ellis, T. Falk, K.A. Olive and Y. Santoso, Nucl.Phys. B652 (2003) 259;
A.  Birkedal-Hansen and B. Nelson, Phys.Rev. D67 (2003) 095006.
%
\bibitem{baryo} See e.g., M. Carena, M. Quiros and C.E.M. Wagner, Nucl. Phys. 
B524 (1998) 3.
%
\bibitem{stop} A. Djouadi, J.L. Kneur, G. Moultaka, Phys.\,Rev.\,Lett.\, 80 
(1998) 1830 
and Nucl.\,Phys.\, B569 (2000) 53; 
A.~Djouadi, Phys. Lett. B435 (1998) 101; 
C. Boehm et al., Phys. Rev. D61 (2000) 095006
and Phys. Rev. D62 (2000) 035012. 
%
\bibitem{qqcd} For the pure QCD corrections to the quark masses, we follow 
the approach of A. Djouadi, M. Spira and P.M. Zerwas, Z. Phys. C70 (1996) 427, 
which has been also used in the program {\tt HDECAY}, except that we work 
in the $\overline{\rm DR}$ scheme. 
%
%
\bibitem{broad} N.~Gray, D.J.~Broadhurst, W.~Grafe and K.~Schilcher,
Z.~Phys.~C48 (1990) 673. 
%
\bibitem{runmass} S.G.\ Gorishny, A.L.\ Kataev, S.A.\ Larin and L.R.\
Surguladze, Mod.\ Phys.\ Lett.\ A5 (1990) 2703; Phys.~Rev.~D43
(1991) 1633.
%
\bibitem{mbdr} L.V. Avdeev and M.Yu. Kalmykov, Nucl. Phys. B502 (1997) 419;
H. Baer, J. Ferrandis, K. Melnikov and X. Tata, Phys. Rev. D66 (2002) 074007.
%
\bibitem{resum}
M.~Carena, D.~Garcia, U.~Nierste and C.~E.~Wagner,
Nucl.\ Phys.\ B577 (2000) 88 [hep-ph/9912516];
%
\bibitem{hrs}
T.~Banks, Nucl.\ Phys.\ B303 (1988) 172;
L.~J.~Hall, R.~Rattazzi and U.~Sarid,
Phys.\ Rev.\ D50 (1994) 7048;
R.~Hempfling, Phys. \ Rev.\ D49 (1994) 6168;
M.~Carena, M.~Olechowski, S.~Pokorski and C.~E.~Wagner,
Nucl.\ Phys.\ B426 (1994) 269;
T. Blazek, S. Pokorski and S. Raby, Phys. Rev. D52 (1995) 4151;  
H. Eberl, K. Hidaka, S. Kraml, W. Majerotto and Y. 
Yamada, Phys. Rev. D62 (2000) 055006.


%
\bibitem{toptev} 
The Tevatron Electroweak Working Group [D0 Collaboration],
hep-ex/0404010.
%
\bibitem{gutthresh} J. Hisano, H. Murayama and T. Yanagida, Nucl. Phys. B402 
(1993) 46, hep-ph/9207279; Y. Yamada, Z. Phys. C60 (1993) 83; 
G.G. Ross and R.G. Roberts, Nucl. Phys. B377 (1992) 571.
%
\bibitem{Martin} Stephen P. Martin, Phys. Rev. D66 (2002) 096001. 
%
\bibitem{CCB} J.A. Casas, A. Lleyda and C. Mu\~noz, Nucl.  Phys. B471 (1996) 3.
%
\bibitem{lemouel}
C. Le Mou\"el, Phys. Rev. D 64 (2001) 075009 and Nucl. Phys. B 607 (2001) 38.
%
\bibitem{Kusenko}  A. Kusenko, P.Langacker and G. Segre, Phys. Rev. D54 (1996) 
5824.
%
\bibitem{sfmix} J. Ellis and S. Rudaz, Phys. Lett. B128 (1983) 248;
M. Drees and K. Hikasa, Phys. Lett. B252 (1990) 127. 
%
\bibitem{egypte} M.M. El Kheishen, A.A. Shafik and A.A. Aboshousha, Phys. Rev. 
D45, 4345 (1992); M. Guchait, Z. Phys. C57, 157 (1993), Erratum: ibid. C61, 
178 (1994).
%
\bibitem{HDECAY} A. Djouadi, J. Kalinowski and M. Spira, Comput. Phys. Commun. 
108 (1998) 56. 
%
\bibitem{fine-tuneold} R. Arnowitt and P. Nath, Phys. Lett. B289 (1992)
368; S. Kelley et al., Nucl. Phys. B398 (1993); Olechowski and S. Pokorski,
Nucl. Phys. B404 (1993) 590; B. de Carlos and J.A. Casas, Nucl. Phys. B309
(1993) 320; G.G. Ross and R.G. Roberts in \cite{gutthresh}.
%
\bibitem{fine-tune} R. Barbieri and G. Giudice, Nucl. Phys. B306 
(1988) 63; for a more recent discussion see, K. Agashe and M. Graesser in Ref.
\cite{agashe}. 
\bibitem{drho0} M.~Veltman, Nucl. Phys. B123 (1977) 89;
A.~Djouadi and C.~Verzegnassi, Phys. Lett. B195 265 (1987);
A.~Djouadi, Nuov. Cim. A100, 357 (1988); 
B.A. Kniehl, Nucl. Phys B347 (1990) 86;
A. Djouadi and P. Gambino, Phys. Rev. D49 (1994) 3499
and Phys. Rev. D51 (1995) 218; 
K.~Chetyrkin, J.~K\"uhn and M.~Steinhauser, Phys. Rev. Lett. 75, 3394 (1995);  
L. Avdeev et al.,  Phys. Lett. B336, 560 (1994); 
G. Degrassi, P. Gambino and A. Vicini, Phys. Lett. B383 (1996) 219;
A. Freitas et al., Nucl. Phys. B632 (2002) 189;
M. Awramik and M. Czakon, Phys. Rev. Lett. 89 (2002) 241801;
M. Awramik, M. Czakon, A. Onishchenko and O. Veretin, Phys. Rev. D68 (2003) 
053004.
%
\bibitem{drhoS} R. Barbieri and L. Maiani, Nucl. Phys. B224, 32 (1983);
C.S. Lim, T. Inami and N. Sakai, Phys. Rev. D29, 1488 (1984);
E. Eliasson, Phys. Lett. 147B, 65 (1984); 
M. Drees and K. Hagiwara, Phys. Rev. D42 (1990) 1709;
M.~Drees, K.~Hagiwara and A.~Yamada, Phys. Rev. D45, 1725, (1992); 
P.~Chankowski, A.~Dabelstein, W.~Hollik, W.~M\"osle, S.~Pokorski and
J.~Rosiek, Nucl. Phys. B417 (1994) 101; D.~Garcia and J.~Sol\`a, Mod.\ Phys.\ 
Lett.\ A 9 (1994) 211. 
%
\bibitem{sloop} A.~Djouadi, P.~Gambino, S.~Heinemeyer, W.~Hollik, C.~J\"unger 
and G.~Weiglein, Phys. Rev. Lett. 78 (1997) 3626 
and Phys. Rev. D57 (1998) 4179. 
%
\bibitem{LEPrho} The LEP Collaborations (ALEPH, L3, DELPHI and OPAL),
hep--ex/0103048. 
\bibitem{BNL} 
Muon $g-2$ Collab., G.W. Bennett et al., Phys. Rev. Lett. 89, 101804
(2002), Erratum--ibid. 89, 129903 (2002), hep--ex/0208001, and
Phys. Rev. Lett. 92, 161802 (2004), hep--ex/0401008.
%
\bibitem{SMgm2} For a recent discussion of theoretical
and experimental developments including the calculation of M.\, 
Knecht and  A.\, Nyffeler, Phys.\, Rev. D65 (2002) 073034, see: M. Davier, S. 
Eidelman, A. Hocker and Z. Zhang, Eur. Phys. J. C27 (2003) 497.
%
\bibitem{g-2old} J. Ellis, J. Hagelin and D.V. Nanopoulos,
Phys. Lett. 116B, 283 (1982); J.A. Grifols and A. Mendez,
Phys. Rev. D26, 1809 (1982); R. Barbieri and L. Maiani,
Phys. Lett. 117B, 203 (1982); D.A. Kosower, L.M. Krauss and N. Sakai,
Phys. Lett. 133B, 305 (1983); U. Chattopadhyay and P. Nath,
Phys. Rev. D53 (1996) 1648; T. Moroi, Phys. Rev. D53 (1996) 6565, and
Erratum: {\it ibid} D56, 4424 (1997); M. Carena, G. Giudice and
C.E. Wagner, Phys.  Lett. B390 (1997) 234.
%
\bibitem{g-2} S.P. Martin and J.D. Wells  Phys. Rev. D64 (2001) 035003.
\bibitem{gm2QED} G. Degrassi and G.F. Giudice, Phys. Rev. D58 (1998) 
053007.
\bibitem{bsg0} S.~Bertolini, F. Borzumati, A. Masiero and G. Ridolfi, 
Nucl. Phys. B353, 591 (1991); R.~Barbieri and G.~Giudice, Phys. Lett. B309, 
86 (1993); F. Borzumati, M. Olechowski and S. Pokorski, Phys. Lett. B349 
(1995) 311.
%
\bibitem{bsg1} 
M. Ciuchini, G. Degrassi, P. Gambino and G.F. Giudice, Nucl. Phys. B534 (1998) 
3; C. Bobeth, M. Misiak and J, Urban, Nucl. Phys. B567 (2000) 153; F. 
Borzumati, C. Greub, T. Hurth and D. Wyler, Phys. Rev. D62 (2000) 075005. 
%
\bibitem{paolo} G. Degrassi, P. Gambino and G.F. Giudice, JHEP 0012 (2000) 
009. 
%
\bibitem{bsg3} M. Ciuchini, G. Degrassi, P. Gambino and G.F. Giudice, 
Nucl. Phys. B527 (1998) 21. 
%
\bibitem{DDK} A. Djouadi, M. Drees and J.L. Kneur, JHEP 0108 (2001) 055.
%
%
\bibitem{benchmark} B.C. Allanach et al., ``The Snowmass points and slopes: 
Benchmarks for SUSY searches", Eur. Phys. J. C25 (2002) 113 [hep-ph/0202233];
N. Ghodbane and H.U. Martyn, hep-ph/0201233.
%
\bibitem{comp} B. Allanach, S. Kraml and W. Porod, ``Comparison of SUSY mass 
spectrum calculations", hep-ph/0207314; JHEP 0303 (2003) 016
[hep-ph/0302102].
\bibitem{compsabi} See the web site: {\tt 
http://cern.ch/kraml/comparison/} 
%

\bibitem{micromegas} G. B\'elanger, F. Boudjema, A. Pukhov and A. Semenov,
``MicrOMEGAs: A Program for calculating the relic density in the MSSM", 
Comput. Phys. Commun. 149 (2002) 103.
%
\bibitem{darksusy}
P. Gondolo, J. Edsjo, L. Bergstrom, P. Ullio and E. Baltz,
DarkSUSY: A Numerical package for dark matter calculations in the MSSM", 
astro-ph/0012234.

\bibitem{nezri} V. Bertin, E. Nezri, J. Orloff, Eur. Phys. J. C26 (2002) 111.

\bibitem{falvard}  A. Falvard, E. Giraud, A.
Jacholkowska, J. Lavalle, E. Nuss, F. Piron, M. Sapinski, P. Salati, R.
Taillet, K. Jedamzik and G. Moultaka, Astropart. Phys. 20 (2004) 467;
see also
E. Giraud {\it et al.} astro-ph/0209230, in {\sl Astronomy, Cosmology and 
Fundamental Physics} proceedings of ESO-CERN-ESA Symposium
eds. P. A. Shaver, L. Di Lella, and A. Gimenez.
%
\bibitem{SDECAY} A. Djouadi, Y. Mambrini and M. Muhlleitner, ``{\tt SDECAY}",
hep--ph/0311167. 
%
\bibitem{SUSYGEN} S. Katsanevas and P. Morawitz, ``SUSYGEN", Comput.\ Phys.\ 
Commun.\  112 (1998) 227; E. Perez, in
``Hamburg 1998/1999, Monte Carlo generators for HERA physics", p.\, 635-651. 
%
\bibitem{SFITTER} R. Lafaye, T. Plehn and D. Zerwas, {\tt SFITTER}, 
hep-ph/0404282.
%
\bibitem{pythia} T. Sjostrand, L. Lonnblad and S. Mrenna, ``PYTHIA", 
hep-ph/0108264.
%
\bibitem{herwig} G.~Corcella {\it et al.}, ``HERWIG 6'', JHEP 0101 (2001) 010. 
%
\end{thebibliography}
\end{document}